\documentstyle[12pt,epsfig]{article}
\newcommand{\be}{\begin{equation}}
\newcommand{\ee}{\end{equation}}
\newcommand{\ba}{\begin{eqnarray}}
\newcommand{\ea}{\end{eqnarray}}

\newcommand{\A}{_{\mbox{\scriptsize A}}}   
\newcommand{\bG}{\bar{\Gamma}} 
\newcommand{\cao}{{\cal O}} 
\newcommand{\dg}{^{\dagger}}
\newcommand{\df}{[\mbox{d}\bar{\psi}\mbox{d}\psi]}
\newcommand{\dfp}{[\mbox{d}\bar{\psi}'\mbox{d}\psi']}
\newcommand{\di}{\mbox{d}\,} 
 
\newcommand{\e}{\mbox{e}}
\newcommand{\f}{{\mbox{\tiny f}}} 
\newcommand{\F}{{\cal F}} 
\newcommand{\g}{\check{G}} 
\newcommand{\ga}{\gamma_5}
\newcommand{\G}{\Gamma} 
\newcommand{\I}{\mbox{\scriptsize I}}   
\newcommand{\II}{\mbox{\scriptsize II}}   
\newcommand{\Id}{\mbox{1\hspace{-0.98mm}l}}   
\newcommand{\la}{\lambda}
\newcommand{\mb}[1]{\quad\mbox{ #1 }\quad}
\newcommand{\na}[1]{\nabla_{#1}}
\newcommand{\p}{{\cal P}}
\newcommand{\pt}{\partial} 
\newcommand{\ra}{\rightarrow}
\newcommand{\re}[1]{(\ref{#1})}
\newcommand{\T}{{\cal T}} 
 
\newcommand{\U}{{\cal U}} 
\newcommand{\V}{\check{V}}

\begin{document}
\renewcommand{\baselinestretch}{1.1} \small\normalsize

\hfill {\sc HU-EP}-00/28

\vspace*{.9cm}

\begin{center}

{\Large \bf Chiral fermions on the lattice and index relations} 

\vspace*{0.8cm}

{\bf Werner Kerler}

\vspace*{0.3cm}

{\sl Institut f\"ur Physik, Humboldt-Universit\"at, D-10115 Berlin, 
Germany}
\hspace{3.6mm}

\end{center}

\vspace*{1.7cm}

\begin{abstract}
Comparing recent lattice results on chiral fermions and old continuum results
for the index puzzling questions arise. To clarify this issue we start
with a critical reconsideration of the results on finite lattices. We then
work out various aspects of the continuum limit. After determining bounds 
and norm convergences we obtain the limit of the anomaly term. 
Collecting our results the index relation of the quantized theory gets 
established. We then compare in detail with the Atiyah-Singer theorem.  
Finally we analyze conventional continuum approaches. 
\end{abstract}

\vspace*{.0cm}

\section{Introduction}

\subsection{General overview} \label{ov}

\hspace{3mm}
The index theorem on finite lattices first appeared in \cite{na94} being
based on the overlap formalism \cite{na93,na95}. With start from the 
Ginsparg-Wilson (GW) relation \cite{gi82} such theorem occurred in
\cite{ha98,lu98}. An explicit form of the massless Dirac operator was found in 
\cite{ne98} from the overlap formalism. The analysis of eigenvalue flows of the
hermitean Wilson-Dirac operator introduced in \cite{na95} initiated a number of
numerical studies of such flows. An exact analytical treatment of these flows 
was presented in \cite{ke99}. The continuum limit of the anomaly term was 
performed for the overlap Dirac operator \cite{ad98,su98} along lines given 
for the Wilson-Dirac operator \cite{ke81,se82} before. Altogether considerable 
progress in various directions has been initiated by these developments.

Comparing lattice results on the index with old continuum results 
\cite{br77,ja77,fu79} puzzling questions arise, which has recently led to 
concern about the relation Tr$\,\ga=0$ \cite{fu99}. These questions are 
related to the observation \cite{ch98} that with the simple GW relation and 
$\ga$-hermiticity the Dirac operator gets normal and the index and the 
corresponding difference at the second real eigenvalue must add up to zero. On 
the other hand, no such restriction is known  with the Atiyah-Singer index 
theorem \cite{at68,at68a}. Since index relations are of fundamental importance
it is highly desirable to clarify this issue. This has been the motivation for
the present paper. To settle the indicated issue one obviously has to care 
about mathematical details. 

To have a reliable basis first a critical reconsideration of the situation on 
finite lattices has been needed. This includes the transformations leading to
the relevant identity (Section 2), the basic relations describing chiral 
properties (Section 3), and a check of GW-related results (Section 4). These 
considerations, an overview of which is given in Section \ref{fin}$\,$, have 
led to side results.

We have taken advantage of the fact that the problem can be dealt with 
considering the theory in a background gauge field. This has allowed us to work
in a unitary space on the finite lattice and in Hilbert space in the limit. 

With respect to the limit several general features had to be clarified (Section
5). To start it has appeared worthwhile to emphasize the precise specification
of the limit. Then firstly it has been shown that the global chiral Ward 
identity -- being the basis of the index relation -- persists in the limit,
which means that Tr$\,\ga=0$ still holds. Secondly it has been checked under 
which conditions the possible emergence of a continuous spectrum does not spoil
the index relation. Thirdly it has been crucial to find minimal conditions on 
the gauge fields and a way to deal with the subtleties of their limit, which we 
have revealed in this context. We give a more detailed overview of these points
in Section \ref{cont}$\,$.

After clarifying the general features mentioned above it has remained to obtain
the limit of the anomaly term (Section 6). With respect to this term 
mathematical questions have been addressed in \cite{ad98,ad00,ad98a,ad00a}. 
However, as we explain in detail in Section \ref{rem}, the very dealing 
with the gauge fields, the use of infinite-series expansions, and the 
drastic assumptions on the gauge fields there are not satisfactory. Therefore 
we had to settle this issue too. Our proper handling of the gauge-fields has 
been based on carefully distinguishing representations of Hilbert space. To 
avoid infinite series we have used ordering relations of operators. An overview
of our approach is found in Section \ref{gind}$\,$.

Having the index relation in the nonperturbative quantized theory established
we have been able to make a precise comparison with the Atiyah-Singer framework
in order to see whether identifications suggested in literature are justified 
(Section 7).  For this purpose it has been necessary to check what the original 
mathematical literature really says and to get the relations relevant for this 
comparison.   It is has turned out that there is indeed a fundamental 
difference. While in nonperturbative quantized theory the chiral subspaces have
the same cardinality and the anomaly stems from chiral noninvariance of the 
action, in the Atiyah-Singer framework the cardinalities of the chiral 
subspaces get different, which is caused by different dimensions of the ones of 
zero modes but equal ones of other modes.  A more detailed overview of this is 
in Section \ref{as}$\,$.

After having the results of nonperturbative quantized theory as well as
knowing its differences to the Atiyah-Singer framework, the question is to
be answered how conventional continuum approaches fit into these schemes
(Section 8).  We consider perturbation theory \cite{ad69,bj69}, the 
Pauli-Villars approach \cite{br77}, and the path-integral method \cite{fu79} 
within this respect. Our analysis shows that these continuum approaches neither
fit in the scheme of the nonperturbative quantized theory nor in that of the 
Atiyah-Singer framework. Instead they rely on a modification of the theory at 
level of the Ward identity. We give a more detailed overview of this in Section 
\ref{ca}$\,$.

We thus finally find that the puzzling question mentioned above resolves in
that one definitely has different situations in nonperturbative quantized 
theory, in the Atiyah- Singer framework, and in conventional continuum 
approaches, respectively, which one must not mix up. Furthermore, it is seen
that the discussions in continuum theory of questions related to the index 
\cite{al85,be96}, which in the past have referred to the path-integral method, 
should more appropriately be based on the nonperturbative definition of the 
quantized theory from the lattice.

\subsection{Basic relations on finite lattice} \label{fin}

\hspace{3mm}
Starting from general Ward identities we see in Section 2 that with a 
background gauge field the operator relations to be investigated are the same 
ones for all expectation values. We introduce a family of alternative chiral 
transformations which lead to the same Ward identity as the usual one, however, allow to transport terms from the action contribution to that of the 
integration measure. This allows us later to discuss tranformations 
\cite{lu98,ch99} recently introduced in connection with the GW relation as well
as an old claim \cite{fu79} that the anomaly would arise from the measure.

In Section 3 we use the resolvent of the Dirac operator $D$ to see that with
normality and $\ga$-hermiticity of $D$ we get rid of eigennilpotents and obtain
chirality of eigenprojections. With this we derive general rules for real 
modes. Applying them to the global chiral Ward identity we get the sum rule for 
chiral differences, the respective difference at eigenvalue zero of $D$ being 
the index$\,$\footnote{This is the usual
definition on the lattice. By the mathematical definition the index is that of 
the Weyl operator associated to the continuum Dirac operator (while the index 
of the latter is zero).}.  It turns out that this sum rule -- since one has to
allow for a nonvanishing index -- generally puts severe restrictions on the 
spectrum of $D$. To give conditions accounting for this we use a particular 
decomposition of $D$ and derive a general expression for appropriate 
one-dimensional spectral constraints.

Turning to the GW relation in Section 4 we make explicit that its general form 
does not guarantee normality of $D$ and point out that the index relation given
in \cite{ha98} has to rely on vanishing of the eigennilpotent in the subspace 
of zero modes only. We see that the simple form of the GW relation is 
actually a spectral constraint to a one-dimensional set. In this special case 
the alternative chiral transformation which transports the anomaly term to the 
measure contribution is seen to get that of \cite{lu98}; we stress that with 
proper zero-mode regularization there still remains the index term in the 
action contribution and thus the action noninvariant.

\subsection{General features in the continuum} \label{cont}

\hspace{3mm}
In Section 5 we first make the definition of the continuum limit precise, as
appears worthwhile in view of some opinions around in literature, and 
emphasize the features important in the present context. 

Since in the limit one can also get a continuous spectrum of $D$ we study the 
respective changes in the spectral representation and investigate their 
consequences for the index relation. It turns out that for a large class of 
spectra, which may be specified by appropriate constraints, still only the 
discrete spectrum contributes to this relation. 

Because the global chiral Ward identity -- on which the index relation is 
based -- is a particular decomposition of Tr$\,\ga=0$ it is of central
importance to establish the validity of this trace relation in the limit. 
We confirm this in several ways and show that it relies on the fact that the 
chiral subspaces of Hilbert space have the same cardinalities. 

We completely avoid the usual reference to classical gauge fields and thus 
prevent the entering of inadequacies as well as of unnecessary restrictions 
as would be introduced by particular ones of them. Instead we determine the 
minimal conditions to be imposed on the lattice gauge fields in order that 
the correct limit exists. This appears also more appropriate in view of future 
considerations of dynamical gauge fields. 

In our more precise examination of the $a\rightarrow0$ limit of the gauge
fields we observe a subtlety which in usual approaches is hidden behind the
notation. It is related to the fact that with $an\rightarrow x$ and fixed
$x$ only the lattice gauge fields at large $|n_{\nu}|$ enter. In order to deal
properly with this fact -- as is a prerequisite for really getting valid
results -- we carefully distinguish $n$-representation and $x$-representation 
of Hilbert space and use the isomorphism between the respective spaces to get
unambiguous norm bounds.

\subsection{Index relation of nonperturbative quantized theory} \label{gind}

\hspace{3mm}
After having the general features of the limit relevant for the global chiral
Ward identity and thus for the index relation investigated, to obtain the
index theorem of nonperturbative quantized theory the limit of the anomaly term
remains to be considered. Though in principle there is a large class of 
appropriate lattice operators, in practice we have to rely on the overlap
Dirac operator \cite{ne98} being the only well-established explicit form 
available. We emphasize that the general form of its spectrum, according to 
what we find in Section \ref{RmR}$\,$, also guarantees that only the discrete 
spectrum contributes to the index relation.

Our determination of the limit of the anomaly term in Section 6 is based on 
norm convergence of operators. In its application we use various site-diagonal 
operators to obtain bounds -- similarly as done for the square of the hermitean 
Wilson-Dirac operator $H$ on the finite lattice in \cite{ne99}. An important 
feature of our approach is that we exploit at many places orderings of 
selfadjoint operators -- which generalizes the getting of lower bounds on 
$H^2$ on the finite lattice in \cite{ne99,he99}. This allows us to avoid any 
infinite-series expansions and their disadvantages.

Performing the $a\rightarrow0$ limit we properly account for the subtlety 
occurring for the gauge fields by carefully distinguishing $n$-representation 
and $x$-representation of Hilbert space and using the isomorphism of the 
respective spaces in dealing with norm bounds.

\subsection{Atiyah-Singer index theorem} \label{as}

\hspace{3mm}
One should be aware of the fact that in the Atiyah-Singer case one considers 
a differential operator and the concept is that of classical fields, while in 
the lattice approach a specific limit of a discrete operator realizes the 
quantum concept. We nevertheless can compare the general structures after
getting the necessary relations in Section 7.

In the Atiyah-Singer framework a nonvanishing index results from different
cardinalities of the chiral subspaces caused by different dimensions of the
chiral subspaces of zero modes and the equality of the dimensions of the 
chiral subspaces of nonzero modes. In contrast to this in nonperturbative 
quantized theory it turns out that the cardinalities of the chiral subspaces
are equal and the difference of the dimensions of the chiral subspaces of 
zero modes has its counterpart in a difference of dimensions of chiral 
subspaces of nonzero real modes. 

In nonperturbative quantized theory the occurrence of a nonvanishing index is 
related to a chirally noninvarinant part of the action which is conceptually 
necessary (to avoid doublers). In the Atiyah-Singer case no such term is
involved. However, by closer inspection we find that there the chiral 
transformation in the topologically nontrivial case does no longer agree 
with the usual one.

We note that in the Atiyah-Singer case obviously the space structure itself 
depends on the particular gauge-field configuration. Fortunately in 
nonperturbative quantized theory, with fixed and equal cardinalities of the 
chiral subspaces, this is not the case.

\subsection{Continuum approaches} \label{ca}

\hspace{3mm}
Having the rigorous relations of the nonperturbative quantized theory as well 
as of the Atiyah-Singer framework, in Section 8 we analyze conventional 
continuum approaches from both points of view. 

We observe that in these approaches on the one hand side there is clearly no 
term in the action producing the anomaly as in nonperturbative quantized 
theory. On the other hand there are also not different cardinalities of chiral 
subspaces as in the Atiyah-Singer framework, which would imply different space 
structures for different gauge configuration and a generalization of the chiral
transformation in the topologically nontrivial case. 

In continuum perturbation theory \cite{ad69,bj69} it is quite obvious that 
instead of the above concepts one performs a modification of the theory at the 
level of the Ward identity.

Reconsidering the Pauli-Villars approach \cite{br77} it turns out that on the 
basis of our general results the usual approximation there is not justified. 
Instead the procedure there is seen to amount to a modification of the theory 
at the level of the Ward identity as one also has in perturbation theory. 

The path-integral method \cite{fu79} starts from redefining an ill-defined
expression. From our results it is obvious that this ill-definedness can 
readily be fixed and is not the actual point. The further analysis then shows 
that this method is essentially a variant of the Pauli-Villars approach so 
that it again amounts to a modification of the theory at the level of the Ward 
identity.

\section{Ward identities} \setcounter{equation}{0}

\subsection{General form of identity} \label{gfm}

\hspace{3mm}
Fermionic Ward identities arise from  the condition that 
$\int \df \e^{-S_\f} \cao$ must not change under a transformation of the 
integration variables. Considering the transformation 
\be
\psi'= \exp(i\eta\G)\psi \mb{,} \bar{\psi}'= \bar{\psi} \exp(i\eta\bar{\G})
\label{tra}
\ee
with parameter $\eta$ this can be expressed by the identity 
\be
\frac{\di}{\di\eta} \int\dfp \e^{-S_\f'} \cao' \Big|_{\eta=0} = 0 
\label{w0s}
\ee
where $S_\f'=\bar{\psi'}M\psi'$. Evaluation of \re{w0s} using the rules for
Grassmann variables gives 
\be
i\int\df \e^{-S_\f}\Big(-\mbox{Tr}(\bar{\G}+\G)\cao 
-\bar{\psi}(\bar{\G}M+M\G)\psi\cao + 
\bar{\psi}\bG\frac{\pt \cao}{\pt \bar{\psi}} -
\frac{\pt \cao}{\pt \psi}\G\psi \Big) = 0 
\label{w1s}
\ee 
(with $\frac{\pt \cao}{\pt \psi}\G\psi \equiv  
\sum_l \frac{\pt \cao}{\pt \psi_l}(\G\psi)_l $). The three contributions
in \re{w1s} stem from the derivative of the integration measure, from that of 
the action, and from that of $\cao$, respectively.

In the present context one usually puts $\cao=1$. We can, however, do better
integrating out the $\psi$ and $\bar{\psi}$ fields in the second term of 
\re{w1s}. For this purpose we use the identity
\ba
0&=&\frac{1}{2}\int\df\Bigg(
\Big(\frac{\pt}{\pt\psi}M^{-1}\Big)_j(\e^{-S_\f}\psi_k\cao) +
\Big(M^{-1}\frac{\pt}{\pt\bar{\psi}}\Big)_k(\e^{-S_\f}
\bar{\psi}_j\cao)\Bigg)
\nonumber \\&=&
\int\df\e^{-S_\f}\Bigg(\bar{\psi}_j\psi_k\cao + 
M^{-1}_{kj}\cao  
+\frac{1}{2}\Big(\frac{\pt \cao}{\pt \psi} M^{-1}\Big)_j \psi_k 
-\frac{1}{2}\bar{\psi}_j\Big(M^{-1} \frac{\pt \cao}{\pt\bar{\psi}}
  \Big)_k \Bigg) 
\label{Ir}
\ea
which follows from the fact that $\int\df(\pt/\pt\psi_l)F=0$ and
$\int\df(\pt/\pt\bar{\psi}_l)F=0$ for any function $F$. Then \re{w1s} becomes
\ba
i\mbox{Tr}\Big(-\bar{\G}-\G + M^{-1}(\bar{\G}M+M\G)\Big) 
\int\df \e^{-S_\f}\cao\nonumber\\+\frac{i}{2}\int\df\e^{-S_\f}\Big(
\frac{\pt \cao}{\pt \psi}M^{-1}R\psi + 
\bar{\psi}RM^{-1}\frac{\pt \cao}{\pt\bar{\psi}} \Big) = 0 
\label{w2}
\ea
where $R = \bar{\G}M-M\G$. To evaluate the derivatives of $\cao$ in \re{w2} 
we note that the fermionic part of $\cao$ with nonzero contribution to the
integral generally is a linear combination of products of type 
${\cal P}= \psi_{j_1}\bar{\psi}_{k_1}\ldots\psi_{j_s}\bar{\psi}_{k_s}$
for which by 
\be 
\int\df \e^{-S_\f}\psi_{j_1}\bar{\psi}_{k_1}\ldots\psi_{j_s}\bar{\psi}_{k_s}= 
\sum_{l_1\ldots l_s}\epsilon^{k_1\ldots k_s}_{l_1\ldots l_s}
            M^{-1}_{j_1 l_1}\ldots M^{-1}_{j_s l_s} \det M 
\label{int}
\ee
(with $\epsilon^{k_1\ldots k_s}_{l_1\ldots l_s}=+1,\,-1$, or $0$ if 
${k_1\ldots k_s}$ even, odd, or no permutation of ${l_1\ldots l_s}$)
we find
\ba
-\int\df\e^{-S_\f} \frac{\pt {\cal P}}{\pt \psi}M^{-1}R\psi =
+\int\df\e^{-S_\f} \bar{\psi}RM^{-1}\frac{\pt {\cal P}}{\pt\bar{\psi}}  =
\qquad\qquad\nonumber\\ \sum_{\mbox{\scriptsize pos}\,z}
\sum_{l_1\ldots l_s}\epsilon^{k_1\ldots k_s}_{l_1\ldots l_s}
M^{-1}_{j_1 l_1}\ldots M^{-1}_{j_z-1 l_z-1}
(M^{-1}RM^{-1})_{j_z l_z} M^{-1}_{j_z+1 l_z+1}
M^{-1}_{j_s l_s} \det M 
\ea
where the first sum is over all positions of the $M^{-1}RM^{-1}$ factor in
the product of the $M^{-1}$. This shows that the terms in \re{w2} with 
derivatives of $\cao$ cancel and we remain with
\be
 i\;\mbox{Tr}\Big(-\bar{\G}-\G + M^{-1}(\bar{\G}M+M\G)\Big) 
 \int\df \e^{-S_\f}\cao =0 \;.
\label{iW}
\ee
 From \re{iW} it is seen that in a background gauge field the expectation 
values factorize so that also for arbitrary $\cao$ (and not only for $\cao=1$) 
it suffices to consider the identity 
\be
 \frac{1}{2}\mbox{Tr}\Big(-\bar{\G}-\G + M^{-1}(\bar{\G}M+M\G)\Big)=0 
\label{W0}
\ee
where $-\frac{1}{2}\mbox{Tr}(\bG+\G)$ is the measure contribution and 
$\frac{1}{2}\mbox{Tr}\Big(M^{-1}(\bar{\G}M+M\G)\Big)$ the action contribution.

\subsection{Chiral transformations} 

\hspace{3mm}
For the global chiral transformation, in which case one has$\,$\footnote{For 
simplicity we write $\ga$ instead of $\ga\bigotimes\Id_s$ wherever this cannot 
cause misunderstandings.} 
\be
\G=\bar{\G}=\ga\;,
\label{t5g}
\ee
the measure contribution vanishes and \re{W0} becomes
\be
\frac{1}{2}\mbox{Tr}(M^{-1}\{\ga,M\})=0 \;.
\label{WM}
\ee 
Obviously this can also be read as Tr$\,\ga=0$, of which the Ward identity is 
the particular decomposition which is dictated by the chiral transformation. 

In order that the Ward identity makes sense we have to care about the existence
of $M^{-1}$. To deal with zero modes of a Dirac operator $D$ we therefore
put $M=D-\zeta$ with the parameter $\zeta$ being in the resolvent set (i.e.~not
in the spectrum of $D$) and let $\zeta$ go to zero only in the final result. 
Thus from \re{WM} we altogether get
\be
\mbox{Tr}\,\ga = \frac{1}{2}\mbox{Tr}\Big((D-\zeta)^{-1}\{\ga,D\}\Big)-
\zeta\,\mbox{Tr}\Big((D-\zeta)^{-1}\ga\Big)=0 \;.
\label{gwa2}
\ee
To have definite names in our discussions we shall call the first term in 
\re{gwa2} anomaly term and the second one index term.

To get the local chiral transformation one has to use 
\be
\G=\bar{\G}=\ga\hat{e}(n) 
\label{t5}
\ee
where $\hat{e}(n)$ is a projection which in lattice-space representation is 
given by
\be
\Big(\hat{e}(n)\Big)_{n''n'}=\delta^4_{n''n}\delta^4_{nn'} \;.
\label{eh}
\ee
Then the identity \re{W0} becomes 
$\frac{1}{2}\mbox{Tr}\Big(M^{-1}\{\ga\hat{e}(n),M\}\Big)=0$,
which may also be read as $\mbox{Tr}\Big(\ga\hat{e}(n)\Big)=0$.
Decomposing $\{\ga\hat{e}(n),M\}$ by
$M=\frac{1}{2}(M-\ga M\ga)+\frac{1}{2}(M+\ga M\ga)$ 
into parts anticommuting and commuting with $\ga$ and inserting 
$M=D-\zeta$ one now has 
\ba
\mbox{Tr}\Big(\ga\hat{e}(n)\Big)&=&
\frac{1}{4}\mbox{Tr}\Big(M^{-1}[\hat{e}(n),[\ga,D]]\Big) +\nonumber\\
&&\frac{1}{4}\mbox{Tr}\Big(M^{-1}\{\hat{e}(n),\{\ga,D\}\}\Big)  -
        \,\zeta\,\mbox{Tr}\Big(M^{-1}\ga\hat{e}(n) \Big) =0\;.
\label{loc}
\ea
The first term in \re{loc} is seen to vanish upon summation over $n$ and 
accordingly corresponds to the divergence of the singlet axial vector current. 
The second term and the third term in \re{loc} are the local versions of the 
anomaly term and of the index term, respectively.

\subsection{Alternative chiral transformations} \label{Alt}

\hspace{3mm}
We introduce a family of alternative global chiral transformation by
\be
\G=\ga-K \mb{,} \bar{\G}=\ga-\bar{K}
\label{tag}
\ee
which inserted into \re{W0} gives 
\be
-\frac{1}{2}\mbox{Tr}(\bG+\G)=+\frac{1}{2}\mbox{Tr}(K+\bar{K})
\label{mes}
\ee
for the measure contribution and 
\be
\frac{1}{2}\mbox{Tr}\Big(M^{-1}(\bG M+M\G)\Big)=
\frac{1}{2}\mbox{Tr}(M^{-1}\{\ga,M\})-\frac{1}{2}\mbox{Tr}(K+\bar{K})
\label{act}
\ee
for the action contribution. Obviously the extra term of the latter cancels 
the measure term so that again the result \re{WM} is obtained for any 
operators $K$ and $\bar{K}$.

While the Ward identity remains the same for these transformations, they
may be used to change the origin of its terms. For example, with
\be
K=\frac{1}{2}M^{-1}\{\ga,D\} \mb{,} \bar{K}=\frac{1}{2}\{\ga,D\}M^{-1} 
\label{tg}
\ee
the anomaly term of \re{gwa2} is transported from the action contribution 
to the measure contribution. Similarly by 
\be
K=\frac{1}{2}M^{-1}\{\ga,M\} \mb{,} \bar{K}=\frac{1}{2}\{\ga,M\}M^{-1} 
\label{tf}
\ee
both terms of \re{gwa2} are transported to the measure contribution. In case
of \re{tf} the action of form $\bar{\psi} M\psi$ is invariant under the 
transformation.

To get the local versions of the alternative chiral transformations one simply 
has to replace $\ga$ of the global cases by $\ga\hat{e}(n)$.  With these 
transformations again the usual result \re{loc} is obtained.

\section{Chiral properties} \setcounter{equation}{0} 

\subsection{Mathematical framework} 

\hspace{3mm}
In numerical work the quantities we are dealing with are given in matrix 
representation. Here we have to care about their mathematical meaning. 
The outcome of the Grassmann integrals can be expressed by determinants,
minors, and generalizations thereof, and in turn by traces of powers of the
operators \cite{ke84}. The definition of such expressions requires 
the mapping to be within the respective space itself. In addition, to have 
independence of the particular basis one must restrict basis transformations 
to similarity transformations as is done in a vector space. 

Thus the theoretical understanding of the quantities in \re{W0} must be that 
of operators in a vector space (with dimension equal to number of sites times 
spinor dimension times gauge-group dimension) which map to this vector space 
itself. The trace then depends solely on the particular operators.
At the same time the formulation of eigenvalue problems -- also important in 
the present context -- becomes possible. 

We need, however, still more. To be able to define adjoint operators we, in 
addition, must introduce an inner product. This means that the vector space 
gets a unitary one. Then basis transformations are restricted to unitary 
ones and one can define normality, including hermiticity and unitarity, 
and also $\ga$-hermiticity of operators (while triangulations of matrices 
and transformation to the Jordan form are no longer generally possible).
We further note that then gauge transformations can be
considered as a particular class of basis transformations.

\subsection{Basic relations}\label{brl} 

\hspace{3mm}
A general operator $D$ has the spectral representation 
$D=\sum_j(\la_j P_j+Q_j)$ with eigenprojections $P_j$ and eigennilpotents 
$Q_j$. Its resolvent is meromorphic, regular at infinity, and given by 
\cite{ka66}
\be
(D-\zeta)^{-1}=-\sum_{j=1}^s\Big((\zeta-\la_j)^{-1}P_j+\sum_{k=1}^{d_j-1}
(\zeta-\la_j)^{-k-1}Q_j^{\,k}\Big)
\label{reso}
\ee 
where $d_j=$ Tr$\,P_j$ is the dimension of the subspace onto which $P_j$ 
projects. For $P_j$ and $Q_j$ one has 
$P_j P_l=\delta_{jl}P_j$, $P_j Q_l=Q_l P_j=\delta_{jl}Q_j$, 
$Q_j Q_l=0$ for $j\ne l$, and $Q_j^{\,d_j}=0$. 

Using hermitean $\gamma$-matrices we require that for the Dirac operator 
$D\dg = \ga D \ga$ holds, i.e.~that it is $\ga$-hermitean$\,$\footnote{With 
anti-hermitean $\gamma$-matrices instead anti-$\ga$-hermiticity 
$\ga D\ga=-D\dg$ is to be required. Our real-mode rules then change into 
purely-imaginary-mode rules.\label{ft}}. Then the resolvent satisfies
\be
(D-\zeta)^{-1}=\ga(D\dg-\zeta)^{-1}\ga\;.
\label{reso1}
\ee
Expressing $(D-\zeta)^{-1}$ and $(D\dg-\zeta)^{-1}$ in \re{reso1} by \re{reso}
and integrating over $\zeta$ around a circle enclosing only one eigenvalue 
$\la_j$ we obtain 
\be
P_j=\ga P_j\dg\ga \mb{for} \la_j \mb{real}
\label{PPd}
\ee
and find that for each complex $\la_j$ with $P_j$ there also occurs a value
$\la_k$ with $P_k$ where $\la_k=\la_j^*$ and $P_k=\ga P_j\dg\ga$. 

To get chiral properties it is necessary that $D$ and $\ga$ have simultaneous 
eigenvectors at least for $\la_j=0$, which means that $P_j$ should commute with
$\ga$. From \re{PPd} we see that
\be
[\ga,P_j]=0 \mb{iff} P_j\dg=P_j \mb{for} \la_j \mb{real}
\label{gaP}
\ee
so that we should have $P_j\dg=P_j$ for this. However, one cannot specify such 
condition itself because $P_j$ would be only available after the eigenvalue 
problem is solved. Therefore, one must find a condition on $D$ which generally 
implies the respective property of $P_j$. A condition which guarantees 
this is $[D,D\dg]=0$, i.e.~normality of $D$\,. Then one gets $P_j\dg=P_j$ for 
all $j$. At the same time it follows that $Q_j=0$ for all $j$ so that the 
spectral representation simplifies to $D=\sum_j\la_j P_j$. A weaker condition
is to require $Q_j=0$ only for $\la_j=0$\,. A realization of this will be 
discussed in Section \ref{fGW}$\,$.

\subsection{Real-mode rules} \label{Rmr}

\hspace{3mm}
With $\ga$-hermiticity and normality of $D$ we can write its spectral 
representation in the form
\be
D=\sum_{j\;(\mbox{\scriptsize Im}\la_j=0)}\la_j P_j^{(5)}
 +\sum_{k\;(\mbox{\scriptsize Im}\la_k>0)}(\la_k P_k^{(1)}+\la_k^* P_k^{(2)})
\label{specd}
\ee
in which the notation of the projections is such that they satisfy
\be
\ga P_j^{(5)}\ga=P_j^{(5)} \mb{,} \ga P_j^{(1)}\ga=P_j^{(2)} \;.
\label{gPg}
\ee 
Because of $[\ga,P_j^{(5)}]=0$ we get $P_j^{(5)}=P_j^{(+)}+P_j^{(-)}$ with
$\ga P_j^{(\pm)}=\pm P_j^{(\pm)}$ so that
\be
\mbox{Tr}(\ga P_j^{(5)})=d_j^{(+)}-d_j^{(-)}
\label{tP5}
\ee
where $d_j^{(\pm)}=$ Tr$\,P_j^{(\pm)}$ is the dimension of the subspace onto
which $P_j^{(\pm)}$ projects. From \re{gPg} according to 
Tr$(\ga P_j^{(1)})=$ Tr$(\ga P_j^{(1)\,2})=$ Tr$(\ga P_j^{(1)}P_j^{(2)})$
we also have  
\be
\mbox{Tr}(\ga P_j^{(1)})=\mbox{Tr}(\ga P_j^{(2)})=0\;.
\label{tP12}
\ee
Since with \re{specd} we get for any function $g(D)$
\be
g(D)=\sum_{j\;(\mbox{\scriptsize Im}\la_j=0)}g(\la_j) P_j^{(5)}
+\sum_{k\;(\mbox{\scriptsize Im}\la_k>0)}\Big(g(\la_k)P_k^{(1)}+g(\la_k^*)
P_k^{(2)}\Big)
\label{gspec}
\ee
using \re{tP5} and \re{tP12} we obtain
\be
\mbox{Tr}(\ga g(D))=\sum_{\la_j \mbox{\scriptsize real}}g(\la_j)I(\la_j)
\label{gD}
\ee
where we have introduced
\be
I(\la_j)=d_j^{(+)}-d_j^{(-)} \mb{for} \la_j \mb{real .}
\label{Idd}
\ee

We now apply \re{gD} to the global chiral Ward identity \re{gwa2}. For the 
index term and the anomaly term we find
\be
   -\lim_{\zeta\rightarrow 0}
\mbox{Tr}\Big((D-\zeta)^{-1}\ga\,\zeta \Big)= I(0)\;,
\label{re0}
\ee
\be
   \lim_{\zeta\ra 0}\,
   \frac{1}{2}\mbox{Tr}\Big((D-\zeta)^{-1}\{\ga,D\}\Big) =
   \sum_{\la_j\ne 0 \mbox{ \scriptsize  real }} I(\la_j)\;,
\label{re1}
\ee
respectively, so that for $\zeta\ra 0$ the identity \re{gwa2} gets the sum rule
for real modes
\be
\mbox{Tr}(\ga)=
\sum_{\la_j\mbox{ \scriptsize real }}I(\la_j) =0 \;.
\label{res}
\ee
 From \re{res} it becomes obvious that one has the same total number of 
right-handed and left-handed modes and that the mechanism leading to a 
nonvanishing index $I(0)$ works via compensating numbers of modes at 
different $\la_j$. Therefore the index can only be nonvanishing if a 
corresponding difference from nonzero eigenvalues exists. Thus in addition 
to 0, allowing for zero modes, there must be at least one further real value 
available in the spectrum. Obviously this puts severe restrictions on the
spectrum of the Dirac operator $D$.

\subsection{Spectral constraints} \label{Con}

\hspace{3mm}
To study how spectral constraints accounting for the restrictions due to
the sum rule \re{res} can be imposed on $D$ we use the decomposition 
\be
 D = u + i v \mb{with}
 u= u\dg= \frac{1}{2} (D+D\dg) \mb{,} v= v\dg = \frac{1}{2i} (D-D\dg)\;.
\label{uv}
\ee
By the $\ga$-hermiticity of $D$ one then in addition has 
\be
u=\frac{1}{2}\ga\{\ga,D\} \mb{,} v=\frac{1}{2}\ga[\ga,D]
\label{uv5}
\ee
from which $[\ga,u]=0$ and $\{\ga,v\}=0$ follow. 

The crucial observation now is that the normality of $D$ implies $\,[u,v]=0$ 
so that for $u$, $v$, and $D$ one gets simultaneous eigenvectors and the 
eigenvalues of $u$ and $v$ are simply the real and imaginary parts, 
respectively, of those of $D$. This opens the way to formulate appropriate
constraints on the spectrum by imposing conditions on $u$ and $v$.

In particular, we may restrict the spectrum to a one-dimensional set specifying
some function $f(u,v)$ and requiring 
\be
f(u,v)=0\;.
\label{f0}
\ee
The function $f(x,y)$ considered for real $x$ and $y$ must be such that for 
$x=0$ and for at least one further value $x=x_l$ the condition $f(x,y)=0$ 
implies $y=0$. Since complex eigenvalues come in pairs we also must have 
$f(x,-y)= f(x,y)$. Requiring that there are $r$ such values $x_l$ we can 
satisfy these conditions by choosing the general form
\be
f(u,v)=u\,(u-x_1)\ldots (u-x_r)\,g(u,v^2) + v^2\,h(u,v^2)
\label{con}
\ee
where $g(u,0)\ne 0$, $h(0,v^2)\ne0$, and $h(x_l,v^2)\ne0$.
With \re{con} the constraint \re{f0} then can be cast into the form of a 
condition on $D$ itself by inserting \re{uv} or \re{uv5}.

A most simple example is obtained putting $r=g=h=1$ and $x_1=2\rho$ in 
\re{con}. Inserting \re{uv5} this is seen to give just the GW-relation of form 
\re{GW}. Similarly the relation $D+\ga D\ga=2a^{2k+1}(\ga D)^{2k+2}$ proposed 
in \cite{fu00} follows from the choice  $r=1$, $x_1=a^{-1}$, 
$g=2a\Big(1+au+(au)^2+\ldots+(au)^{2k}\Big)$, and 
$h=2a\sum_{l=0}^k \left(\begin{array}{l}k+1\\l+1\\\end{array}\right)
(av)^{2l} (au)^{2(k-l)}$. In these special cases because
of $r=1$ the sum rule \re{res} gets simply $I(0)+ I(x_1) =0$. Apparently
it is straightforward to construct further constraints along these lines.

\section{Discussion of GW-related results} \setcounter{equation}{0}

\subsection{Form and meaning of GW-relation}\label{fGW} 

\hspace{3mm}
The general GW relation \cite{gi82} can be written as
\be
\{\ga,D\}= 2 D \ga R D 
\label{aGW}
\ee
where $R$ is a hermitean operator which is trivial in Dirac space. From 
\re{aGW} using $\ga$-hermititcity of $D$ and $[\ga,R]=0$ one obtains
$[D,D\dg]=2D\dg[R,D]D\dg$. Thus it is seen that one should have $[R,D]=0$ in 
order that $D$ gets normal. To get this property generally one has to put
$R$ equal to a multiple of the identity. 

To check what happens for general $R$ we insert \re{aGW} into the identity 
\re{gwa2} and find
\be
\mbox{Tr}\,\ga = 
\mbox{Tr}(\ga RD)-\zeta\,\mbox{Tr}\Big(\ga(D-\zeta)^{-1}\Big)
                 +\zeta^2\,\mbox{Tr}\Big(\ga R(D-\zeta)^{-1}\Big)=0\;.
\label{e2}
\ee
Dividing \re{e2} by $\zeta$, expressing $(D-\zeta)^{-1}$ by \re{reso}, and 
integrating over $\zeta$ around a circle enclosing only the eigenvalue 
$\la_k=0$ we obtain 
\be
\mbox{Tr}\,\ga = 
\mbox{Tr}(\ga RD)+\mbox{Tr}\Big(\ga(P_k+RQ_k)\Big)=0 \mb{for} \la_k=0\;.
\label{e2i}
\ee 

We can evaluate this a little further considering the set of eigenvectors with
$Df_{kl}=0$ and $l=1,\ldots,g_k$. Using $D+D\dg=2D\dg RD$, as one has from 
\re{aGW} and $\ga$-hermititcity, we obtain $[\ga,D]f_{kl}=0$. Thus the set of 
$f_{kl}$ can be chosen such that $\ga f_{kl}=s_{kl}f_{kl}$ with $s_{kl}=\pm1$ 
and one gets $g_k=g_k^{(+)}+g_k^{(-)}$ where $g_k^{(\pm)}$ denotes the numbers 
of modes with $s_{kl}=\pm1$, respectively. It is important to note here that 
the $f_{kl}$ span the geometric eigenspace $M_k'$ of $D$ with dimension $g_k$ 
while $P_k$ projects onto the algebraic eigenspace $M_k$ with dimension $d_k$, 
and that $g_k\le d_k$. For $g_k< d_k$ we may decompose $P_k$ as 
$P_k=P_k'+P_k''$ correponding to $M_k=M_k'\bigoplus M_k''$. Then 
$\mbox{Tr}(\ga P_k)$ in \re{e2i} can be written as 
$g_k^{(+)}-g_k^{(-)}+\mbox{Tr}(\ga P_k'')$. Comparing with the relation given 
in \cite{ha98} it is seen that the term $\mbox{Tr}\Big(\ga(P_k''+RQ_k)\Big)$ 
is missing there. Because of $d_k-g_k=\mbox{Tr }P_k''=\mbox{rank }Q_k$ the
vanishing of this term requires $g_k=d_k$. As discussed in Section \ref{brl} 
this would hold for all $Q_j$ if $D$ would we normal, which is, however, not 
the case for general $R$. Since $Q_k=0$ is only needed for $\la_k=0$, actually 
only this weaker condition is to be shown. This is indeed rigorously possible 
\cite{ke00} by working out detailed properties of nilpotents. 

Putting $R=(2\rho)^{-1}\Id$ in \re{aGW} to have normality of $D$ we arrive at 
the simple GW relation
\be
\{\ga,D\}= \rho^{-1} D \ga D \;.
\label{GW}
\ee
Requiring also $\ga$-hermiticity of $D$, the condition \re{GW} means that 
$\rho(D+D\dg)=D D\dg=D\dg D$ should hold, i.e.~that $D/\rho-1$ should be 
unitary. Thus the actual content of \re{GW} is the restriction of the spectrum 
of $D$ to the circle through zero with center at $\rho$. In fact, as already 
mentioned, it is simply a spectral constraint of type \re{f0} with \re{con}. 

For $R=(2\rho)^{-1}\Id$ the form \re{e2} of the identity \re{gwa2} becomes 
\be
\mbox{Tr}\,\ga = 
(2\rho)^{-1}\mbox{Tr}(\ga D)-\zeta\,\mbox{Tr}\Big(\ga(D-\zeta)^{-1}\Big)
            +(2\rho)^{-1}\zeta^2\,\mbox{Tr}\Big(\ga (D-\zeta)^{-1}\Big)=0\;.
\label{e2e}
\ee
In \re{e2e}, because of $Q_j=0$ for normal $D$, the term with $\zeta^2$ does no
longer contribute in the limit $\zeta\rightarrow 0$ so that one remains with
the form 
\be
(2\rho)^{-1}\mbox{Tr}(\ga D)
\label{an}
\ee 
of the anomaly term, while the index term is the same as in \re{gwa2}.
As for all constraints with $r=1$ in \re{con} the sum 
rule \re{res} now has only two terms 
\be
\mbox{Tr}\,\ga = I(0)+I(2\rho)=0 \;. 
\label{ress}
\ee
With the special form \re{an} of the anomaly term one can also use the relation 
Tr$(\ga D)= \sum_{\la_j\ne 0 \mbox{ \scriptsize real }} \la_j\,I(\la_j)$
following from \re{gD} to obtain $(2\rho)^{-1}\mbox{Tr}(\ga D)= I(2\rho)$.

\subsection{Alternative transformations in GW case} 

\hspace{3mm}
For the alternative transformation with \re{tg}, which transports the anomaly 
term to the measure contribution, in case of the GW relation \re{GW} with the 
form \re{an} of the anomaly term one gets 
\be
K=(2\rho)^{-1}\ga D \mb{,} \bar{K}=(2\rho)^{-1}D\ga \;.
\label{tl}
\ee
This gives the transformation introduced in infinitesimal form in \cite{lu98}
which by \re{GW} leaves the classical action $\bar{\psi} D\psi$ invariant. 
With \re{tl} the measure contribution gets $(2\rho)^{-1}\mbox{Tr}(\ga D)$. 
However, there still remains the index term 
$-\zeta\mbox{Tr}(\ga(D-\zeta)^{-1}$) of the Ward identity in the action 
contribution. 

The remaining action contribution is missing in \cite{lu98} since no zero-mode 
regularization has been used. Thus it looks there like the action would also
be invariant in the quantum case with zero modes, as is not correct. In an
independent step, which without motivation uses a decomposition of Tr$\,\ga=0$
equivalent to \re{e2e}, what should have been in the action contribution is 
calculated there from the measure term. 

The generalized transformation of \cite{ch99} is obtained by putting 
$K=\ga SD$, $\bar{K}=DT\ga$ in \re{tag} where $S$ and $T$ are hermitean
operators which are trivial in Dirac space. If $D$ satisfies the general
GW relation \re{aGW} with $R=S+T$ the classical action $\bar{\psi} D\psi$ 
is invariant under this transformation.

\section{Continuum limit} \setcounter{equation}{0} 

\subsection{Steps towards the limit} \label{Lim}

\hspace{3mm}
The continuum limit is the final part of the nonperturbative definition of a 
quantum field theory provided by the lattice formulation. It appears worthwhile
to remind of the general scheme and of the succession of the steps: 
(1) The quantization prescription is to discretize a continuum action such 
that the classical continuum limit, namely that of the action alone, 
gives the continuum expression. 
(2) With the discretized action the functional integrals are evaluated on 
the finite lattice; the lattice-spacing parameter drops out at this stage. 
(3) The infinite-volume limit is taken on the basisis of a sequence of 
finite-lattice results.
(4) Letting the parameters approach a critical value the limit of zero lattice 
spacing is obtained.

Since for the present purpose it suffices to consider the quantum field theory 
of fermions in a background gauge field we consider here the details 
for this case. Obviously it remains to deal with Steps (3) and (4). By Step (3)
our unitary space gets of infinite dimension. In order to be able to perform 
limits we must complete it to a separable Hilbert space. In the relations 
considered so far the main change is that the spectrum of the operator $D$ then
can include a continuum. Further the existence of traces is to be checked.

Step (4) is most familiar in theories with a bare mass which is parametrized as 
$m_{\mbox{\scriptsize bar}}(a)=a\,m_{\mbox{\scriptsize physical}}$ such that
with decreasing lattice spacing $a$ it goes to a critical singularity (where
the correlation lenght diverges). In this way, given the $n$-representation of 
space, letting $an\ra x$, one gets the $x$-representation. 

In the present context Step (4) has to deal with the gauge field $U_{\mu n}$. 
The situation is similar to that for a theory with 
$m_{\mbox{\scriptsize bar}}(a)\ra0$ in that we have to parametrize such that 
$U_{\mu n}(a)\rightarrow\Id$ for $a\rightarrow 0$.  However, unlike a bare mass
the field $U_{\mu n}$ depends on $n$. This introduces a subtlety, frequently 
hidden in the notation, which is revealed carefully distinguishing 
$n$-representation and $x$-representation.

The mostly used relation of lattice gauge fields $U_{\mu n}$ to continuum 
gauge fields $A_{\mu}(x)$ then reads $U_{\mu n}=
{\cal P}\exp\Big(\int_{an}^{a(n+\hat{\mu})}\mbox{d}^4 x\,A_{\mu}(x)\Big)$ 
where ${\cal P}$ denotes path ordering. Now $B_{\mu n}$ in 
$U_{\mu n}=\exp(iB_{\mu n})$ is a certain average over $aA_{\mu}(y)$ where
$an_{\nu}\le y_{\nu}\le a(n_{\nu}+1)$ and for small $a$ one essentially has 
$B_{\mu n}\approx a A_{\mu}(an)$. In the limit requiring $an\rightarrow x$ 
for $a\ra0$ with fixed $x$ implies that also $|n_{\nu}|\rightarrow\infty$. 
The latter not only needs Step (3) as a prerequisite but also causes the
subtlety that this are only the $U_{\mu n}$ with large $|n_{\nu}|$ which enter.

Starting from the lattice the gauge fields are given there. Therefore we intend
to find the precise conditions on the lattice gauge fields which are needed in 
order that the correct limit exists (which is in contrast to the frequent view 
of prescribing classical fields). 

After taking the infinite-volume limit we are in $n$-representation which
means that the abstract Hilbert space is realized as that of sequences. 
Alternatively we can realize it as that of square-integrable functions.
This is what we have in $x$-representation which we prefer after taking
the limit $a\ra0$ of vanishing lattice spacing. Since the space of sequences
as well as the space of square-integrable functions both represent a
separable Hilbert space they are isomorphic so that we can use both 
descriptions.

\subsection{Spectrum in Hilbert space} \label{Spc}

\hspace{3mm}
In Hilbert space the spectrum of $D$ can also get continuous parts. Using the 
decomposition \re{uv} we then can rely on the spectral representations of the 
selfadjoint operators $u$ and $v$ in terms of operator Stieltjes integrals
\be
D=\int \di E_{\alpha}^{\I}\:\alpha + i\int \di E_{\beta}^{\II}\:\beta \;.
\label{Dab}
\ee
Because $D$ is normal the spectral families of $u$ and $v$ commute, 
$[E_{\alpha}^{\I},E_{\beta}^{\II}]=0$. From 
$\ga$-hermiticity, inserting \re{Dab} into $D=\ga D\dg\ga$, it follows that 
\be
\ga E_{\alpha}^{\I}\ga=E_{\alpha}^{\I}\mb{,}
\ga E_{\beta}^{\II}\ga=E_{-\beta}^{\II}\;.
\label{E5}
\ee
Using $\int \di E_{\alpha}^{\I}=\Id$ and $\int \di E_{\beta}^{\II}=\Id$ 
we can write \re{Dab} as
\be 
D=\int \di E_{\alpha}^{\I} \int \di E_{\beta}^{\II}\:(\alpha+i\beta) \;.
\label{Dabs}
\ee
In order to decompose this with respect to values with $\beta=0$, $\beta>0$, 
and $\beta<0$ we define 
\be
E_{\alpha}^{(5)}=E_{\alpha}^{\I} \int_{-0}^{+0}\di E_{\beta}^{\II} \;,\quad
E_{\beta}^{(1)}=\left\{\begin{array}{ll} E_{\beta}^{\II} &\mbox{for }\beta>0\\
0 &\mbox{otherwise} \end{array}\right. \;,\quad
E_{\beta}^{(2)}=\left\{\begin{array}{ll} E_{\beta}^{\II} &\mbox{for }\beta<0\\
0 &\mbox{otherwise} \end{array}\right. 
\label{E12}
\ee
for which by \re{E5} 
\be
\ga E_{\alpha}^{(5)}\ga=E_{\alpha}^{(5)}\mb{,} \ga E_{\beta}^{(1)}\ga=E_{\beta}^{(2)}
\label{g5E}
\ee
holds. With \re{E12} now \re{Dabs} becomes 
\be
D=\int \di E_{\alpha}^{(5)} \alpha  
 +\int \di E_{\alpha}^{\I} \int \di E_{\beta}^{(1)} (\alpha+i\beta)
 +\int \di E_{\alpha}^{\I} \int \di E_{\beta}^{(2)} (\alpha+i\beta)
\label{specc}
\ee
which generalizes \re{specd}. For functions $g(D)$ we thus get 
\be
g(D)=\int \di E_{\alpha}^{(5)} g(\alpha)  
 +\int \di E_{\alpha}^{\I} \int \di E_{\beta}^{(1)} g(\alpha+i\beta)
 +\int \di E_{\alpha}^{\I} \int \di E_{\beta}^{(2)} g(\alpha+i\beta)\;
\label{gspecc}
\ee
where we require $g(\la)$ to be continuous to guarantee that the sum of the 
Stieltjes integrals still equals 
$g(D)=\int \di E_{\alpha}^{\I} \int \di E_{\beta}^{\II}\:g(\alpha+i\beta)$.
With the relation \re{gspecc} we now have the generalization of \re{gspec}.

\subsection{Modification of real-mode relations} \label{RmR}

\hspace{3mm}
With the spectral representation \re{gspecc} $\mbox{Tr}(\ga g(D))$ is defined 
in terms of the traces of the spectral family. Since by \re{E5} one has 
$[\ga,E_{\alpha}^{\I}]=0$ and with
\re{g5E} gets $\ga E_{\beta}^{(1)}=\ga E_{\beta}^{(1)\,2}=
E_{\beta}^{(2)}\ga E_{\beta}^{(1)}$ and $\ga E_{\beta}^{(2)}=
E_{\beta}^{(1)}\ga E_{\beta}^{(2)}$ it follows that only the first integral in 
\re{gspecc} contibutes and we remain with
\be
\mbox{Tr}\Big(\ga g(D)\Big)=\int\di\!\Big(\mbox{Tr}(\ga E_{\alpha}^{(5)})\Big)
  \;g(\alpha)\;.  
\label{gpm}
\ee
Because from \re{g5E} we have $[\ga,E_{\alpha}^{(5)}]=0$ we can decompose 
$E_{\alpha}^{(5)}$ as $E_{\alpha}^{(5)}=E_{\alpha}^{(+)}+E_{\alpha}^{(-)}$ with 
$\ga E_{\alpha}^{(\pm)}=\pm E_{\alpha}^{(\pm)}$ so that \re{gpm} becomes
\be
\mbox{Tr}(\ga g(D))=
\int\di\!\Big(\mbox{Tr}(E_{\alpha}^{(+)}-E_{\alpha}^{(-)})\Big)\;g(\alpha)\;.
\label{gDg}
\ee
Further, in \re{gDg} we can separate discrete and continuous parts of the 
spectum putting $E_{\alpha}^{(\pm)}=
E_{\alpha}^{\mbox{\scriptsize d}(\pm)}+E_{\alpha}^{\mbox{\scriptsize c}(\pm)}$.
With $\int 
\di\!\Big(\mbox{Tr}\,E_{\alpha}^{\mbox{\scriptsize d}(\pm)}\Big)\;g(\alpha)=
\sum_jg(\alpha_j)d_j^{(\pm)}$
and $I(\alpha_j)=d_j^{(+)}-d_j^{(-)}$ this gives
\be
\mbox{Tr}(\ga g(D))=\sum_j g(\alpha_j)I(\alpha_j)
+\int \di\!\Big(\mbox{Tr}(E_{\alpha}^{\mbox{\scriptsize c}(+)}-
                     E_{\alpha}^{\mbox{\scriptsize c}(-)})\Big)\;g(\alpha) 
\label{gDc}
\ee
which generalizes \re{gD}.

An important observation is now that for the large class of spectra with only 
discrete points on the real axis still only the discrete spectrum contributes 
to \re{gDc}. This follows because $E_{\alpha}^{\mbox{\scriptsize c}(\pm)}$ then
must be constant outside the respective points and, on the other hand, by 
definition is continuous.

It should be noted that the indicated class includes the cases where the
spectrum is restricted to curves which cross the real axis. This occurs,
for example, with the circle in the simple GW case. Generally this can be
achieved by imposing appropriate constraints as discussed in Section 
\ref{Con}$\,$.

\subsection{Tr$\,\ga$ in Hilbert space} \label{tr5}

\hspace{3mm}
The chiral Ward identities are particular decompositions of Tr$\,\ga$ in the 
global case and of Tr$\,\ga\hat{e}(n)$ in the local one. That these traces are 
zero makes the respective relations identities. We have to check if this is 
also guaranteed in the infinite-volume limit.

In the local case this clearly holds since $\ga\hat{e}(n)$ is restricted to
a subspace of finite dimension. In the global case we have, firstly, to be 
consistent with the fact that the infinite-lattice result is to be considered
as limit of the results on finite lattices with increasing size, which for
Tr$\,\ga$ means that we get a sequence with all members zero and therefore 
also zero in the limit. Secondly, consistency is to be required with the fact
that summing the local result over $n$ we get the global one which with 
respect to Tr$\,\ga$ again amounts to a sequence with all members and thus 
also the limit being zero. Thirdly, with $\{\gamma_{\mu},\ga\}=0$ and 
$\gamma_{\mu}^2=1$ one in the usual way gets $\mbox{Tr}\,\ga=-\mbox{Tr}\,\ga$ 
and can hardly escape to conclude that $\mbox{Tr}\,\ga=0$. Thus not to run into
severe contradictions we must still have the relation $\mbox{Tr}\,\ga=0$ in 
the limit.

We now consider what this in more detail means in Hilbert space. For this
purpose we first remember that Tr$\,\ga$ is actually a shorthand for 
Tr$(\ga\bigotimes \Id_s)$, or introducing $\Gamma_5= \ga\bigotimes\Id_s$, of 
Tr$\;\Gamma_5=0$. We also note that the projection involved in the local case is
of form $\tilde{P}_j=\Id_{\gamma}\bigotimes \tilde{p_j}$. In this form we are 
free to choose, more generally, any set of orthogonal projections $\tilde{p_j}$ 
which project onto a subspaces of finite dimension. With $\sum_{j=1}^{\infty} 
\tilde{p_j}=\Id_s$ we then have $\sum_{j=1}^{\infty} \tilde{P}_j=\Id$. Now 
clearly Tr$(\Gamma_5\sum_{j=1}^{N}\tilde{P}_j)=0$ holds for finite $N$. Thus 
we can define Tr$\,\Gamma_5$ by the limit of the sequence of trace class 
operators $\Gamma_5\sum_{j=1}^{N}\tilde{P}_j$ as
\be
\mbox{Tr}(\Gamma_5)=\lim_{N\ra\infty}\mbox{Tr}
(\Gamma_5 \sum_{j=1}^{N}\tilde{P}_j)
\label{5l}
\ee
or more simply, performing the traces first, by 
\be
\mbox{Tr}(\Gamma_5)=\sum_{j=1}^{\infty}\mbox{Tr} (\Gamma_5 \tilde{P}_j)\;.
\label{5ls}
\ee
Both of these regularizations give Tr$\,\Gamma_5=0\,$. It should be also noted
that the structure of \re{5ls} is analogous to the definition of the trace in 
\re{gpm} in terms of the spectral representation \re{gspecc}.

The relation Tr$\,\Gamma_5=0$ actually expresses the fact that there is no 
asymmetry of the chiral subspaces. To make this explicit we decompose 
$\tilde{P}_j$ as $\tilde{P}_j= \tilde{P}_j^{(+)}-\tilde{P}_j^{(-)}$ where 
$\tilde{P}_j^{(\pm)}=\frac{1}{2}(1\pm\ga)\bigotimes \tilde{p_j}$. We then have 
$\Gamma_5 \tilde{P}_j=\tilde{P}_j^{(+)}-\tilde{P}_j^{(-)}$ and 
\be
\mbox{Tr}\Big(\Gamma_5 \tilde{P}_j\Big)=
\mbox{Tr}\,\tilde{P}_j^{(+)}-\mbox{Tr}\,\tilde{P}_j^{(-)}=0 
\label{atr}
\ee
which means that the subspaces onto which $\tilde{P}_j^{(+)}$ and 
$\tilde{P}_j^{(-)}$ project have the same dimension. Since the total space
is made up of such pairs of chiral subspaces it is seen that, with 
$\tilde{P}^{(\pm)}=\sum_{j=1}^{\infty}\tilde{P}_j^{(\pm)}$, the operators
$\tilde{P}^{(+)}$ and $\tilde{P}^{(-)}$ project onto spaces which have the 
same cardinality. Summing up \re{atr} we get again Tr$\,\Gamma_5=0$. 

Mathematically another definition of the equality of cardinalities is that by 
the existence of an appropriate bijective mapping \cite{ch77}. 
This is provided here by $\Gamma_4= \gamma_4\bigotimes\Id_s$ with which one 
gets $\Gamma_4 \tilde{P}^{(\pm)}_j\Gamma_4 =\tilde{P}^{(\mp)}_j$ as well as
$\Gamma_4 \tilde{P}^{(\pm)}\Gamma_4 =\tilde{P}^{(\mp)}$.

\subsection{Limit $a\ra0$ of gauge fields} \label{nxlim}

\hspace{3mm}
The operators $\U_{\mu}$ in Hilbert space which involve the gauge fields
in $n$-representation are given by
\be
(\U_{\mu})_{n'n}=U_{\mu n}\delta^4_{n',n+\hat{\mu}} \;.
\label{U}
\ee
With them we get four unitary plaquette operators defined by
$\p^{(1)}_{\mu\nu}=\U_{\mu}\dg\U_{\nu}\dg\U_{\mu}\U_{\nu}$ and
\be
\p^{(2)}_{\mu\nu}=\U_{\nu}\p^{(1)}_{\mu\nu}\U_{\nu}\dg\,,\; 
\p^{(3)}_{\mu\nu}=\U_{\mu}\p^{(2)}_{\mu\nu}\U_{\mu}\dg\,,\; 
\p^{(4)}_{\mu\nu}=\U_{\nu}\dg\p^{(3)}_{\mu\nu}\U_{\nu}\,,\; 
\p^{(1)}_{\mu\nu}=\U_{\mu}\dg\p^{(4)}_{\mu\nu}\U_{\mu}
\label{UpU}
\ee
in terms of which we can introduce the field operators
\be
\F^{(\alpha)}_{\mu\nu}=\Id-\p^{(\alpha)\dag}_{\mu\nu}
\label{F}
\ee
which have the advantageous property of being site diagonal
\be
\Big(\F^{(\alpha)}_{\mu\nu}\Big)_{n'n}=-i\delta^4_{n'n}F^{(\alpha)}_{\mu\nu,n}
\;.\label{pF}
\ee
The $F^{(\alpha)}_{\mu\nu,n}$ depend in an obvious way only on the 
$U_{\mu n}$. Further, using the unitary operators $\T_{\mu}$ which in 
$n$-representation are given by
\be
(\T_{\mu})_{n'n}=\delta^4_{n',n+\hat{\mu}} 
\ee
one gets the site-diagonal combinations 
$(\T_{\mu}\dg \U_{\mu})_{n'n}=\delta^4_{n'n}U_{\mu n}$ and
\be
(\T_{\la}\dg \F^{(\alpha)}_{\mu\nu}\T_{\la})_{n'n}=-i\delta^4_{n'n}
        F^{(\alpha)}_{\mu\nu,n+\hat{\la}}\mb{,}
(\T_{\la} \F^{(\alpha)}_{\mu\nu}\T_{\la}\dg)_{n'n}=-i\delta^4_{n'n}
        F^{(\alpha)}_{\mu\nu,n-\hat{\la}}
\label{TFT}
\ee
which will be also needed in the following. 

With $U_{\mu n}=\exp(iB_{\mu n})$ we now list the minimal conditions on the 
$B_{\mu n}$ which will be seen to be necessary in the limit. Firstly we require
\be
B_{\mu n}/a\rightarrow A_{\mu} \mb{and} an\rightarrow x \mb{for} 
a\rightarrow 0
\label{Bax}
\ee 
which, as already stressed in Section \ref{Lim}$\,$, 
implies that at the same time $n\ra\infty\,$. By \re{Bax} the values $A_{\mu}$ 
and $x$ are related which constitutes a pointwise definition of the matrix 
function $A_{\mu}(x)\,$. Secondly 
\be
(B_{\mu,n+\hat{\nu}}-B_{\mu n})/a^2\ra b_{\mu\nu} \mb{and} an\ra x \mb{for} 
a\ra 0
\label{Baa+}
\ee 
is to be required for $\nu\ne\mu$. The values 
$b_{\mu\nu}$ and $x$ in \re{Baa+} then provide a pointwise 
definition of the matrix function $\pt A_{\mu}(x)/\pt x_{\nu}$ with 
$\nu\ne\mu$. In this way in addition $A_{\mu}(x)$ gets continuous in 
$\nu$-direction. Thirdly we must have 
\be
(B_{\mu,n}-B_{\mu,n\mp\hat{\mu}})/a\ra 0\mb{and} an\ra x \mb{for}a\ra 0\;,
\label{Baa0}
\ee
which provides continuity of $A_{\mu}(x)$ also in $\mu$-direction. Fourthly we 
need 
\be
(B_{\mu,n}-B_{\mu,n-\hat{\mu}}-B_{\mu,n\mp\hat{\la}}+
B_{\mu,n-\hat{\mu}-\mp\hat{\la}})/a^2 \ra 0\mb{and} an\ra x \mb{for}a\ra 0    
\label{Baa2}
\ee
for $\nu\ne\mu$ giving continuity of $\pt A_{\mu}(x)/\pt x_{\nu}$ in all 
directions.

Using the explicit form of $F^{(\alpha)}_{\mu\nu,n}$ and the
Baker-Campbell-Hausdorff formula one finds that with these conditions 
one gets independently of $\alpha$
\be
F^{(\alpha)}_{\mu\nu,n}/a^2\ra F_{\mu\nu} \mb{and}an\ra x \mb{for}a\ra0 
\label{Fax}
\ee
where the values $F_{\mu\nu}$ and $x$ provide the pointwise definition of the 
function $F_{\mu\nu}(x)$. On the other hand, $F_{\mu\nu}(x)$ then is in the
usual way given by the functions $A_{\mu}(x)$ and $\pt A_{\mu}(x)/\pt x_{\nu}$
which follow here from their pointwise definitions. Thus $F_{\mu\nu}(x)$ is 
also continuous. In addition these conditions guarantee that one has also
\be
F^{(\alpha)}_{\mu\nu,n\pm\hat{\la}}/a^2\ra F_{\mu\nu} \mb{and}an\ra x 
\mb{for}a\ra0
\label{Faxt}
\ee
for the translated fields of \re{TFT}, which we shall need too. Conversely, 
in order that \re{Fax} and \re{Faxt} hold all of the indicated conditions are 
necessary in full.

\subsection{Gauge-field norms} \label{gfn}

\hspace{3mm}
In the following we shall need a appropriate behaviors of gauge-field 
norms. We therefore in addition require the quantities $A_{\mu}$ and 
$b_{\mu\nu}$ in \re{Bax} and \re{Baa+} to be bounded, which implies 
boundedness of $A_{\mu}(x)$ and $F_{\mu\nu}(x)$.

With \re{pF} the norm squared of the operators $\F^{(\alpha)}_{\mu\nu}$ gets 
in $n$-representation 
\be
||\F^{(\alpha)}_{\mu\nu}||^2=\sup_{\phi}\sum_{n}\phi_n\dg 
F^{(\alpha)\dag}_{\mu\nu,n}F^{(\alpha)}_{\mu\nu,n}\phi_n 
\ee
where $\sum_{n}\phi_n\dg \phi_n=1\,$. In the limit $a\ra0$ with \re{Fax} we 
obtain in $x$-representation 
\be
||\F^{(\alpha)}_{\mu\nu}||^2\rightarrow a^4\sup_\phi\int\mbox{d}^4x\,
  \phi\dg(x) F_{\mu\nu}(x)^2\phi(x)
\label{Fl}
\ee
with $\int\mbox{d}^4x\,\phi(x)\dg \phi(x)=1\,$. Thus, since $F_{\mu\nu}(x)$ is 
bounded, the norm vanishes in the limit. 

Further, for the norm squared of the site diagonal combinations 
$\T_{\mu}\dg \U_{\mu}-\Id\,$ we have 
\be
||\T_{\mu}\dg \U_{\mu}-\Id||^2=\sup_{\phi}\sum_{n}\phi_n\dg 
(2-U_{\mu n}-U_{\mu n}\dg)\phi_n \;.
\ee
In the limit with \re{Bax} we now obtain in $x$-representation 
\be
||\T_{\mu}\dg \U_{\mu}-\Id||^2\ra a^2\sup_{\phi}\int\mbox{d}^4x\,
  \phi\dg(x)A_{\mu}(x)^2\phi(x)
\label{TU}
\ee
which with the boundedness of $A_{\mu}(x)$ also vanishes for $a\ra0$.

\section{Limit of index relation} \label{ilim}\setcounter{equation}{0} 

\subsection{Form of anomaly term} \label{olim}

\hspace{3mm}
To calculate the limit $a\rightarrow0$ of the anomaly term in the index
relation we have to choose an explicit lattice form of $D$. In principle there
is a large class of appropriate ones with the same limit out of which we could 
take one. In practice, however, the only well-established explicit form 
available is the overlap Dirac operator \cite{ne98}
\be
D=\rho\,\Big(1+\ga \epsilon(H)\Big)  
\label{neu0}
\ee
where $\epsilon(H)$ is the sign function and $H$ the hermitean Wilson-Dirac 
operator. The operator \re{neu0} is normal and $\ga$-hermitean. It satisfies 
the GW relation \re{GW} and in addition to zero allows for the real eigenvalue
$2\rho$. We also note that with this operator by \re{gDc} only the discrete 
spectrum contributes to the index relation. Thus it is a suitable choice.

Since \re{neu0} satisfies \re{GW} the index term simplifies to the form 
\re{an} which by inserting \re{neu0} becomes 
\be
\frac{1}{2}\mbox{Tr}(\epsilon(H))\;.
\label{Iro}
\ee
Obviously $\rho$ has dropped out in \re{Iro} so that the particular value of 
$\rho\ne0$ in \re{neu0} is irrelevant in the present context. Of course, to
get a finite result in \re{Iro} $\epsilon(H)$ should belong to the trace
class. We shall come back to this at the end of Section \ref{alim}$\,$.

For $H$ in \re{neu0} we use the hermitean Wilson-Dirac operator of form
\be
H=\ga\Big(\sum_{\mu}(\gamma_{\mu}D_{\mu}+W_{\mu})-1\Big)
\label{H}
\ee
in which the Wilson parameter is $r=1$ and where for $m$ the favorable value
\cite{ne99} $m=-1$ has been chosen. In \re{H} we have 
\be
D_{\mu}=\frac{1}{2}(\U_{\mu}\dg-\U_{\mu})\mb{,}
W_{\mu}=\Id-\frac{1}{2}(\U_{\mu}\dg+\U_{\mu})
\label{D,W} 
\ee
the $\U_{\mu}$ being given by \re{U}.

Since $H$ is a selfadjoint operator in Hilbert space we can express the index
term \re{Iro} as
\be
\frac{1}{2}\mbox{Tr}(\epsilon(H))=\frac{1}{2\pi}\int_{-\infty}^{+\infty}
\di s\,\mbox{Tr}\frac{H}{H^2+s^2}\;. 
\label{sH}
\ee
Decomposing $H^2$ into parts with and without $\gamma$-matrices gives $H^2=L-V$ 
with 
\be
L=\Id+\sum_{\mu\ne\nu}W_{\mu}W_{\nu}\;,
\ee                                           
\be
V=\sum_{\mu\ne\nu}\Big(\frac{1}{2} \gamma_{\mu}\gamma_{\nu}[D_{\mu},D_{\nu}]
+\gamma_{\mu}[D_{\mu},W_{\nu}]\Big) \;.
\ee                                           
Because at least four $\gamma$-matrices are needed to contribute to the trace 
we can write
\be
\mbox{Tr}\frac{H}{H^2+s^2}=\mbox{Tr}\Big(HGVGVG(\Id+V\frac{1}{L+s^2-V})\Big)
\label{TrV}
\ee
where $G=(L+s^2)^{-1}$.

\subsection{Limit of anomaly term} \label{alim}

\hspace{3mm}
The task now is to perform the $a\ra0$ limit of \re{sH} with \re{TrV}.
To get rid of the last term in \re{TrV} we show that we have the norm 
convergence $\Id+V(L+s^2-V)^{-1}\Rightarrow\Id$ for $a\ra0$, which requires
that that $||V(L+s^2-V)^{-1}||\rightarrow0$. According to 
$||V(L+s^2-V)^{-1}||\le||V||\,||(L+s^2-V)^{-1}||$ this follows if 
$||V||\rightarrow0$ and if $(L+s^2-V)^{-1}$ is bounded. To derive this 
boundedness we first split off from $L$ a positive part, 
$L=1+L_++L_F$, where with $\na{\mu}=\Id-\U_{\mu}$ 
\be
L_+=\frac{1}{8}\sum_{\mu\ne\nu}\Big(\na{\mu}\na{\nu}( \na{\mu}\na{\nu})\dg
                      +\na{\mu}\dg\na{\nu}( \na{\mu}\dg\na{\nu})\dg\Big)\;,
\label{L+}
\ee
\be
L_F=-\frac{1}{8}\sum_{\mu\ne\nu}\Big(
\U_{\nu}(\F_{\mu\nu}^{(1)}+\F_{\mu\nu}^{(4)\dag})+
(\U_{\nu}(\F_{\mu\nu}^{(1)}+\F_{\mu\nu}^{(4)\dag}))\dg \Big) \;.
\label{LF}
\ee
We then note that due to the selfadjointness of the operators we obtain the 
ordering relation  $(L+s^2-V)\ge 1+L_F-V\ge 1-||L_F||-||V||$. 
It implies $ (L+s^2-V)^{-1}\le 1-||L_F||-||V||$ from which one gets 
$||(L+s^2-V)^{-1}||\le 1-||L_F||-||V||$. Thus all we need is that 
$||L_F||\rightarrow0$ and $||V||\rightarrow0$. To deal with $V$ we note that
\ba
4[D_{\mu},D_{\nu}]=
 \U_{\mu}\U_{\nu}\F^{(1)}_{\mu\nu}+\U_{\mu}\dg\U_{\nu}\dg\F^{(3)}_{\mu\nu}+
 \U_{\nu}\U_{\mu}\dg\F^{(4)}_{\mu\nu}+\U_{\nu}\dg\U_{\mu}\F^{(2)}_{\mu\nu}\;, 
\nonumber\\
4[D_{\mu},W_{\nu}]=
 \U_{\mu}\U_{\nu}\F^{(1)}_{\mu\nu}-\U_{\mu}\dg\U_{\nu}\dg\F^{(3)}_{\mu\nu}+
 \U_{\nu}\U_{\mu}\dg\F^{(4)}_{\mu\nu}-\U_{\nu}\dg\U_{\mu}\F^{(2)}_{\mu\nu}\;.
\label{DDW} 
\ea
 From \re{LF} and \re{DDW} it becomes obvious that $||L_F||\rightarrow0$ and 
$||V||\rightarrow0$ follow from $||\F^{(\alpha)}_{\mu\nu}||\ra0$, which we 
have by \re{Fl}. In \re{TrV} we thus remain with $\mbox{Tr}\Big(HGVGVG\Big)$ 
and after evaluation of the $\gamma$-traces get 
\ba
\mbox{Tr}\frac{H}{H^2+s^2}\ra\sum_{\mu\nu\la\tau\sigma}
\epsilon_{\mu\nu\la\tau}\mbox{Tr}
\Big(G(W_{\sigma}-1/4)G[D_{\mu},D_{\nu}]G[D_{\la},D_{\tau}]
\nonumber\\ +2GD_{\mu}G[D_{\nu},W_{\sigma}]G[D_{\la},D_{\tau}]
+2G[D_{\nu},W_{\sigma}]GD_{\mu}G[D_{\la},D_{\tau}]\Big)\;.
\label{TrDW}
\ea

To evaluate the commutators in \re{TrDW} we use that all 
$F^{(\alpha)}_{\mu\nu,n}$ by \re{Fax} lead to the same limit so that in
\re{DDW} we can replace all $\F^{(\alpha)}_{\mu\nu}$ by one of them. Further 
the $\U_{\mu}$ factors there can be replaced by the $\T_{\mu}$ since 
$\U_{\mu}\Rightarrow\T_{\mu}$ according to \re{TU}. Thus instead of \re{DDW} 
we now have
\be
4[D_{\mu},D_{\nu}]\Rightarrow
 (\T_{\mu}+\T_{\mu}\dg)(\T_{\nu}+\T_{\nu}\dg)\F^{(1)}_{\mu\nu}\;,\quad
4[D_{\mu},W_{\nu}]\Rightarrow 
 (\T_{\mu}+\T_{\mu}\dg)(\T_{\nu}-\T_{\nu}\dg)\F^{(1)}_{\mu\nu}\;.
\label{DDF} 
\ee
In \re{TrDW} we further can replace $G$ by $G^{(0)}=(L^{(0)}+s^2)^{-1}$ where 
$L^{(0)}=\Id+\sum_{\mu\ne\nu}W_{\mu}^{(0)}W_{\nu}^{(0)}$ with $W_{\mu}^{(0)}=
\Id-\frac{1}{2}(\T_{\mu}\dg+\T_{\mu})$ since $G\Rightarrow G^{(0)}$, as follows
from $G=G^{(0)}(\Id-(L-L^{(0)})G)$ because of $||(L-L^{(0)})||\ra 0$ and  
boundedness of $G$. This boundedness is obtained from the ordering relation 
$(L+s^2)\ge 1+L_F\ge 1-||L_F||$ which implies $||(L+s^2)^{-1}||\le 1-||L_F||$.
That $||(L-L^{(0)})||\ra 0$ follows from \re{TU} noting that 
$||\U_{\mu}\U_{\nu}-\T_{\mu}\T_{\nu}||\le||\T_{\mu}\dg\U_{\mu}-\Id||+
||\T_{\nu}\dg\U_{\nu}-\Id||+||\T_{\mu}\dg\U_{\mu}-\Id|| 
||\T_{\nu}\dg\U_{\nu}-\Id||$.
Because of $||(\U_{\mu}\dg\mp\U_{\mu})-(\T_{\mu}\dg\mp\T_{\mu})||\le
2||\T_{\mu}\dg\U_{\mu}-\Id||$ and \re{TU} we can also replace 
$GD_{\mu}G$ by $G^{(0)2}(\T_{\mu}\dg-T_{\mu})/2$ and $G(1-W_{\sigma})G$ by 
$G^{(0)2}(\T_{\sigma}\dg+T_{\sigma})/2$ in \re{TrDW}. Further, because of 
\re{TFT} with \re{Faxt}, we can interchange there $\F^{(1)}_{\mu\nu}$ with the 
$\T_{\la}$ and the $\T_{\la}\dg$, and also with $G^{(0)}$ which is a function 
of such operators. In addition we make use of the fact that 
$\F^{(1)}_{\mu\nu}$ and $-\F^{(1)}_{\nu\mu}$ have the same limit. 

With the mentioned replacements and interchanges \re{TrDW} becomes
\be
G^{(0)3}\sum_{\mu\nu\la\tau}
\epsilon_{\mu\nu\la\tau}\mbox{Tr}\Bigg(
\sum_{\sigma}\Big(\frac{(\T_{\sigma}\dg-\T_{\sigma})^2}
{2(\T_{\sigma}\dg+\T_{\sigma})}+\frac{3}{4}-
\frac{\T_{\sigma}\dg+\T_{\sigma}}{2}\Big) 
\prod_{\rho}\frac{\T_{\rho}\dg+\T_{\rho}}{2}
\F^{(1)}_{\mu\nu}\F^{(1)}_{\la\tau} \Bigg)\;.
\label{TrDW1}
\ee
Remembering $L=1+L_++L_F$ with \re{L+} and \re{LF} we see that $L_0\ge\Id$.
Thus inserting \re{TrDW1} into \re{sH} we can perform the integral 
$\int\!\mbox{d} s\,G^{(0)3}$ and arrive at
\be
\frac{1}{2}\mbox{Tr}\,\epsilon(H)\ra \sum_{\mu\nu\la\tau}
\epsilon_{\mu\nu\la\tau}\mbox{Tr}
\Big({\cal C}\F^{(1)}_{\mu\nu}\F^{(1)}_{\la\tau}\Big)
\ee
where
\be
{\cal C}=\frac{3}{16}\sum_{\sigma}\Bigg(\frac{(\T_{\sigma}\dg-\T_{\sigma})^2}
{2(\T_{\sigma}\dg+\T_{\sigma})}+\frac{3}{4}-
\frac{\T_{\sigma}\dg+\T_{\sigma}}{2}\Bigg) 
\prod_{\rho}\frac{\T_{\rho}\dg+\T_{\rho}}{2} \frac{1}{(\sqrt{L^{(0)}}\,)^5}\;.
\label{cnn0}
\ee
With \re{pF} we then have 
$\mbox{Tr}({\cal C}\F^{(1)}_{\mu\nu}\F^{(1)}_{\la\tau})
=-\sum_n {\cal C}_{nn}\mbox{tr}(F^{(1)}_{\mu\nu,n}F^{(1)}_{\la\tau,n})$. 
Since ${\cal C}$ depends on the $\T_{\mu}$ only, ${\cal C}_{nn}$ is a number 
independent of $n$. We emphasize that this is the crucial point which
becomes clear here by properly distinguishing $n$-representation and 
$x$-representation. 

The indicated number can be found analytically utilizing the representation
\be
{\cal C}_{nn}=-\frac{3}{16}
\prod_{\rho}\Big(\frac{1}{2\pi}\int_{-\pi}^{\pi}\di k_{\rho} \cos k_{\rho}\Big)
\frac{1+\sum_{\sigma}(\cos k_{\sigma}-1+
\sin^2k_{\sigma}/\cos k_{\sigma})}{\Big(\sqrt{1+\sum_{\mu\ne\nu}
(1-\cos k_{\mu})(1-\cos k_{\nu})}\;\Big)^5}\;.
\label{cnn}
\ee 
The integral in \re{cnn} can be evaluated, using an identity analogous to that
introduced in \cite{ka81}, as done in \cite{ad98,su98} and carefully worked out
in \cite{su98}, which gives ${\cal C}_{nn}= -(32\pi^2)^{-1}$. 

Inserting \re{Fax} we now have for $a\ra0$ in $x$-representation
\be
\frac{1}{2}\mbox{Tr}\,\epsilon(H)\ra \frac{1}{32\pi^2}
\int \mbox{d}^4x\sum_{\mu\nu\la\tau}
\epsilon_{\mu\nu\la\tau}\mbox{ tr}\Big(F_{\mu\nu}(x)F_{\la\tau}(x)\Big)\;.
\label{FF}
\ee
While before taking the trace we have needed only boundedness of 
$F_{\mu\nu}(x)$ in our derivations, to get a finite result in \re{FF} it should
decrease sufficiently fast at infinity. This simply means that to our 
conditions on the gauge fields introduced in Section \ref{nxlim} we have to add
the slightly stronger one that $F_{\mu\nu}(x)$  should be such that \re{FF} 
remains finite$\,$\footnote{Alternatively one may admit that the index also can
get infinite. Then, of course, it remains to prescribe how infinity should be 
approached.}. 

In terms of operator properties the effect of this condition is to make 
$\epsilon(H)$ a trace-class operator. As a side remark we note that while this 
holds for $\epsilon(H)$ itself, it does not necessarily apply to other 
operators by which $\epsilon(H)$ is possibly expressed. For example, using the 
relations of Sections \ref{Rmr} and \ref{RmR} one may express Tr$\,\epsilon(H)$
as a trace of a difference of projection operators which individually are not 
trace-class. If one wishes one can also properly deal with this formulation
using operator regularizations of the types discussed in Sections 
\ref{tr5}$\,$, \ref{arel}$\,$, and \ref{direl}$\,$. 

A further observation is that in \re{FF} modifications of the gauge fields on 
a set of measure zero do not matter. We therefore can considerable relax our 
conditions on the gauge fields requiring them only almost everywhere and thus
admit much more general classes of fields.

\subsection{Index theorem} \label{qlim}

\hspace{3mm}
With the Dirac operator \re{neu0} by \re{gDc} only the discrete spectrum 
contributes to the identity \re{gwa2}. The index term then by \re{re0} is 
$I(0)$ as defined in \re{Idd}. The anomaly term for the operator \re{neu0} 
simplifies to \re{an} and gets the form \re{Iro} which has the limit \re{FF}. 
We thus obtain
\be
\mbox{Tr}\,\ga= I(0) +\frac{1}{32\pi^2} \int \mbox{d}^4x\sum_{\mu\nu\la\tau}
\epsilon_{\mu\nu\la\tau}\mbox{ tr}\Big(F_{\mu\nu}(x)F_{\la\tau}(x)\Big)=0 
\label{ind}
\ee
which is the index theorem as it follows in the framework of nonperturbative 
quantized theory on the basis of the chiral Ward identity. 

Since $I(0)$ takes integer values only this must also hold for the gauge-field 
term in \re{ind}. With respect to the topological side of this we remember that
the minimal conditions on the gauge fields needed for the limit of the anomaly
term have precisely lead to continuity of $F_{\mu\nu}(x)$. Because continuity 
is a prerequisite for homotopy classes and thus for topological invariants this
hints at the underlying mechanism.

\subsection{Remarks on literature} \label{rem}

\hspace{3mm}

In \cite{fu98} the path-integral method \cite{fu79} has been adapted to the 
limit of the index relation on the lattice. This procedure not only disregards 
the specific origin of the anomaly on the lattice but also runs into the same
problems as in the continuum which are described in Section \ref{compi}$\,$. 
Further, a zero mode regularization is missing in \cite{fu98} so that, 
similarly as in \cite{lu98}, the relation as a whole is not really obtained. 
Nevertheless, in this context the important observation has been made 
\cite{fu99} that one has to care about the relation Tr$\,\ga=0\,$. In contrast 
to the view in \cite{fu99}, however, from our rigorous considerations here 
it becomes clear that in the limit this relation still holds.

To establish the limit of the index relation as a whole, in addition to 
showing that Tr$\,\ga=0$ still holds, we also had to check under which 
conditions a possible continuous spectrum does not spoil the relation, which 
has not been done in literature. 

In the other papers which occurred solely the limit of the anomaly term is
considered for the overlap Dirac operator.  In \cite{su98} mathematical details
are not addressed. In \cite{ad98} the $a\rightarrow0$ limit is performed on the
finite lattice, which is actually not possible, and incorrect factorizations of
square roots are involved. In \cite{ad00} a somewhat modified approach has 
mainly only been announced. In \cite{ad98a,ad00a} the respective formulations 
have improved$\,$\footnote{An email of the present author, pointing out a list 
inadequacies, as well as the availability of the present work in hep-lat have 
obviously helped within this respect.}. However, the very treatment of the 
gauge fields, the use of infinite-series expansions, and the drastic 
assumptions on the gauge fields are still unsatisfactory. 

A crucial point concerning the gauge fields is that the $a\rightarrow0$ limit 
drives $|n_{\nu}|$ to infinity so that only the lattice fields at the 
respective limiting values contribute. We are able to handle this subtlety 
properly by carefully distinguishing $n$-representation and $x$-representation 
of Hilbert space and exploiting the isomorphism of the related spaces in the 
calculations of norms (on the basis of details explained in Sections \ref{Lim}
and \ref{gfn}).

In literature this subtlety remains hidden behind the notation. However, 
independently of the particular formulation it is there and its consideration
is a prerequisite for a true definition of the limit and for a proper account
of the occurring norms. In view of this the respective approaches cannot be
really valid.

Distinguishing representations has also allowed us to give the proper 
justification of the evaluation of \re{cnn0}, being that ${\cal C}_{nn}$ is 
a number independent of $n$, which in usual formulations cannot be seen and
has not been noticed before. 

In our derivations we extensively use splittings into two terms and exploit 
ordering relations for self-adjoint operators to get bounds of one of these 
terms and to show its vanishing in the limit. This is clearly not only much 
more efficient than infinite-series expansions but also applies to more
general cases where infinite series do not converge. 

In contrast to literature we neither refer to classical gauge fields nor to
gauges but determine the minimal conditions to be imposed on the lattice 
gauge fields in order that the correct limit exists. In this way we avoid 
unnecessary restrictions as well as inadequate conditions which would be
introduced by prescribing particular ones of such fields. 

The difficulties in \cite{ad98,ad00,ad98a,ad00a} are related to several 
shortcomings there. Firstly, by not accounting for the subtlety of the 
gauge-field limit described above proper track of the bounds in the limit is 
lost, which spoils the general validity of the approach. Secondly, the use of 
infinite-series expansions is not only technically inefficient but also 
unnecessarily excludes large classes of fields. Thirdly, the reference to 
particular classical gauge fields introduces unnecessary 
complications. Fourthly, the drastic assumptions on the gauge fields (as 
vanishing outside a bounded region, being in a particular gauge, and more) 
admit only a small fraction of the cases of interest.

The remarks in \cite{ad98a} with respect to our performing of the trace are
obviously rooted in the difficulties of the approach there and not in any 
study of details of the present work. From Section \ref{alim} it can be seen 
that we get firm results in the Hilbert-space framework which are not affected 
in any way by the speculations on classical fields, gauges, and singularities 
in \cite{ad98a}.

\section{Comparison with Atiyah-Singer theorem} \label{CompA}
\setcounter{equation}{0} 

\subsection{Form of theorem} \label{frel}

\hspace{3mm}
The Atiyah-Singer theorem for elliptic differential operators on compact
manifolds without boundaries relates the analytical index of such operators to 
topological invariants. The relevant papers$\,$\footnote{Further papers contain 
different types of proofs or present various generalizations. Other ones, also 
occasionally referred to, actually consider manifolds with boundaries.} are 
\cite{at68,at68a}. Ref.~\cite{at68} gives the proof in K-theory (i.e.~entirely 
within algebraic topology) and Ref.~\cite{at68a} mainly translates the results 
to cohomology. The special case of the Dirac operator, to be considered here, 
is treated in Section 5 of \cite{at68a}. 

In the case of interest here the theorem says that the index of the Weyl 
operator $D^{(+)}$ associated to the Dirac operator $D\A$ equals the Pontryagin 
index, which we may write as 
\be
\mbox{index }D^{(+)}=-\frac{1}{32\pi^2}
\int \mbox{d}^4x\sum_{\mu\nu\la\tau}
\epsilon_{\mu\nu\la\tau}\mbox{ tr}\Big(F_{\mu\nu}(x)F_{\la\tau}(x)\Big)
\label{indA}
\ee
where index $D^{(+)}=\mbox{dim ker }D^{(+)}-\mbox{dim ker }D^{(+)\dag}$. The 
operator $D^{(+)}$ maps between two spinor spaces, i.e.~from $E^{(+)}$ to 
$E^{(-)}$ say, and its adjoint $D^{(+)\dag}\equiv D^{(-)}$ back \cite{at68a}.

\subsection{Analytical relations} \label{arel}

\hspace{3mm}
For our comparison we need to know what in detail holds for the analytical 
index. To get the connection to zero modes one has to note that 
$D^{(-)}D^{(+)}$ maps within $E^{(+)}$ and that $D^{(+)}D^{(-)}$ maps
within $E^{(-)}$.  Both of these operators are selfadjoint and nonnegative. 
Since one is on a compact manifold their spectra are discrete and the 
degeneracies for nonzero eigenvalues are finite. Except for zero modes one has 
the same spectra for both operators. This follows since, firstly, with the 
eigenequation $D^{(-)}D^{(+)}\varphi_j=\kappa_j\varphi_j$ in $E^{(+)}$ 
multiplying by $D^{(+)}$ one gets the eigenequation 
$D^{(+)}D^{(-)}(D^{(+)}\varphi_j)=\kappa_j(D^{(+)}\varphi_j)$ in $E^{(-)}$. 
Secondly, in case of degeneracy at an eigenvalue $\kappa_j$ considering 
$\langle D^{(+)}\varphi_{j r'}|D^{(+)}\varphi_{j r}\rangle=
\langle\varphi_{j r'}|D^{(-)}D^{(+)}\varphi_{j r}\rangle= 
\kappa_j\langle\varphi_{j r'}|\varphi_{j r}\rangle$ one sees that the 
eigenspaces must have the same dimension exept for $\kappa_j=0$. 

For $\kappa_j>0$ we thus have dim$\,E^{(+)}_j=$ dim$\,E^{(-)}_j$ or in terms 
of eigenprojections
\be
\mbox{Tr}\Big(P^{(+)}_j-P^{(-)}_j\Big)=0 \mb{for} \kappa_j>0\;.
\label{i1}
\ee
To see what happens for $\kappa_j=0$ one has to note that 
ker$\,D^{(+)}=$ ker$\,D^{(-)}D^{(+)}$ and 
ker$\,D^{(-)}=$ ker$\,D^{(+)}D^{(-)}$, which
from left to right is obvious and in the opposite direction follows from 
$\langle\varphi|D^{(-)}D^{(+)}\varphi\rangle=
\langle D^{(+)}\varphi|D^{(+)}\varphi\rangle$. 
We therefore for $\kappa_j=0$ get dim$\,E^{(\pm)}_j=$ dim~ker$\,D^{(\pm)}$ and 
in terms of eigenprojections 
\be 
\mbox{index}\,D^{(+)}= \mbox{Tr}\Big(P^{(+)}_j-P^{(-)}_j\Big)
                       \mb{for} \kappa_j=0\;.
\label{i0}
\ee
We thus make the remarkable observation that for a nonvanishing index of 
$D^{(+)}$ the dimensions of the eigenspaces $E^{(+)}_j$ and $E^{(-)}_j$ for 
$\kappa_j=0$ are different while for $\kappa_j>0$ they are always equal. This 
means that for a nonvanishing index the spaces $E^{(+)}$ and $E^{(-)}$ have 
different cardinalities.

Defining $P^{(\pm)}=\sum_j P^{(\pm)}_j$ the projections $P^{(+)}$ and $P^{(-)}$
are the ones projecting onto the spaces $E^{(+)}$ and $E^{(-)}$, respectively.
Combining \re{i1} and \re{i0} one formally gets $\mbox{index}\,D^{(+)}= 
\mbox{Tr}(P^{(+)}-P^{(-)})$. This can be defined as the limit of a sequence of 
trace-class operators $\sum_{j=1}^{N}(P^{(+)}_j-P^{(-)}_j)$ by
\be
\mbox{index}\,D^{(+)}= 
\lim_{N\ra\infty}\mbox{Tr}\sum_{j=1}^{N}\Big(P^{(+)}_j-P^{(-)}_j\Big)
\label{ig}
\ee
or more simply, performing the traces first, by  
\be
\mbox{index}\,D^{(+)}= 
\sum_{j=1}^{\infty}\mbox{Tr} \Big(P^{(+)}_j-P^{(-)}_j\Big)\;.
\label{igs}
\ee
Obviously the regularizations in \re{ig} and \re{igs} are analougous to 
those in \re{5l} and \re{5ls}, however, the space structure here is 
fundamentally different.

\subsection{Dirac operator} \label{direl}

\hspace{3mm}
The Dirac operator $D\A$ is defined as the composition of the two maps 
$D^{(+)}$ and $D^{(-)}$ \cite{at68a}. Thus in the combined space 
$E=E^{(+)}\bigoplus E^{(-)}$ introducing 
\be
\hat{D}^{(\pm)}=\left\{\begin{array}{ll}
D^{(\pm)}&\mbox{ for mapping from } E^{(\pm)} \mbox{ to } E^{(\mp)} \\
0& \mbox{ otherwise }\\\end{array}\right.\;.
\label{dh}
\ee
we have $D\A=\hat{D}^{(+)}+\hat{D}^{(-)}$. Obviously $D\A$ is selfadjoint. 
From \re{dh} one obtains 
$D\A^2=\hat{D}^{(+)}\hat{D}^{(-)}+\hat{D}^{(-)}\hat{D}^{(+)}$. 
This connects to the eigenequations considered for the operators 
$D^{(+)}D^{(-)}$ and $D^{(-)}D^{(+)}$ before and leads to the spectral 
representation
\be
D\A=\sum_j\la_j(P^{(+)}_j+P^{(-)}_j) \mb{with} \la_j^2=\kappa_j \;.
\label{di}
\ee

Introducing a function $f(D\A^2)$, \re{i1} generalizes to 
\be
\mbox{Tr}\Big(f(D\A^2)(P^{(+)}_j-P^{(-)}_j)\Big)=0 \mb{for} \kappa_j>0
\label{i1f}
\ee
and, in addition requiring $f(0)=1$, \re{i0} to 
\be 
\mbox{index}\,D^{(+)}= \mbox{Tr}\Big(f(D\A^2)(P^{(+)}_j-P^{(-)}_j)\Big)
                       \mb{for} \kappa_j=0\;.
\label{i0f}
\ee
Because $D\A^2$ is nonnegative one can choose
$f(D\A^2)$ such that $f(D\A^2)P^{(\pm)}$ get trace-class operators. This 
provides a convenient regularization with which one can readily combine 
\re{i1f} and \re{i0f} to 
\be
\mbox{index}\,D^{(+)}= 
\mbox{Tr}\Big(f(D\A^2)(P^{(+)}-P^{(-)})\Big)\;.
\label{Pf}
\ee
This regularization is, for example, used with $f(D\A^2)=\exp(-t D\A^2)$ 
in an alternative proof of the Atiyah-Singer theorem based on the heat equation
\cite{at73}. While technically the regularization in \re{Pf} has some advantage
it clearly does not affect the structures of the spaces $E^{(+)}$ and $E^{(-)}$
in any way.

We note that $D\A$ given by \re{dh} anticommutes with $P^{(+)}-P^{(-)}\,$,
\be
\{(P^{(+)}-P^{(-)})\,,D\A\}=0\;.
\label{5h}
\ee
We further find that the transformation \re{tra} with 
\be
\G=\bar{\G}=P^{(+)}-P^{(-)}
\label{t5ga}
\ee
leaves the classical action $\bar{\psi} D\A\psi$ invariant. Obviously
this is the global chiral transformation in the Atiyah-Singer case. 

Though the Atiyah-Singer framework is a classical one we may compare the 
structures inserting its global chiral transformation with \re{t5ga} into the 
identity \re{W0} of the quantum case. We thus get
\be
-\mbox{Tr}\Big(P^{(+)}-P^{(-)}\Big)-
\zeta\mbox{Tr}\Big((D\A-\zeta)^{-1}(P^{(+)}-P^{(-)})\Big)=0
\label{gwa3}
\ee
which is made up of the measure contribution and of the index term of the 
action contribution while the anomaly term in the latter by \re{5h} vanishes.

\subsection{Relations with $\ga$} \label{grel}

\hspace{3mm}
In physical applications the connection of the relations in Sections \ref{arel}
and \ref{direl} to ones with $\ga$ is of interest. To make the respective 
details explicit we consider the projections $P^{(\pm)}_j$ onto the occurring 
eigenspaces. We are are free to express these operators in the forms
$P^{(+)}_j=\left(\begin{array}{ll}p^{(+)}_j&0\\0&0\\\end{array}\right)$ and
$P^{(-)}_j=\left(\begin{array}{ll}0\;\;&0\!\\0\;\;&p^{(-)}_j\!\\
\end{array}\right)$~. Introducing $\ga=\left(\begin{array}{rr}1&0\\0&-1\\
\end{array}\right)$ this can be written as
\be
P^{(\pm)}_j=\frac{1}{2}(1\pm\ga)\bigotimes p^{(\pm)}_j\;.
\label{i1g}
\ee
For $\kappa_j>0$ from \re{i1} we have Tr$\,P^{(+)}_j=$ Tr$\,P^{(-)}_j$ which 
is satisfied if Tr$\,p^{(+)}_j=$ Tr$\,p^{(-)}_j$. Thus identifying $p^{(+)}_j=
p^{(-)}_j=p_j$ in \re{i1g} we get a valid representation of $P^{(\pm)}_j$ and
obtain
\be 
P^{(+)}_j-P^{(-)}_j=\ga\bigotimes p_j \mb{for} \kappa_j>0 \;.
\label{P5}
\ee
For $\kappa_j=0$ according to \re{i0} we must allow for different values of 
Tr$\,P^{(+)}_j$ and Tr$\,P^{(-)}_j$. Then if Tr$\,P^{(+)}_j>$ Tr$\,P^{(-)}_j$ 
we need Tr$\,p^{(+)}_j>$ Tr$\,p^{(-)}_j$. To realize this we identify the space
onto which $p^{(-)}_j$ projects with a subspace of that onto which $p^{(+)}_j$ 
projects so that $p^{(+)}_j p^{(-)}_j=p^{(-)}_j p^{(+)}_j=p^{(-)}_j$. The 
latter allows to decompose $p^{(+)}_j$ as $p^{(+)}_j= 
p^{(-)}_j+(p^{(+)}_j-p^{(-)}_j)$ into orthogonal parts and inserting this 
decomposition into \re{i1g} gives $P^{(+)}_j-P^{(-)}_j= 
\ga\bigotimes p^{(-)}_j+\frac{1}{2}(1+\ga)\bigotimes(p^{(+)}_j- p^{(-)}_j)$.
Proceeding analogously for Tr$\,P^{(+)}_j<$ Tr$\,P^{(-)}_j$ we altogether have
\be
P^{(+)}_j-P^{(-)}_j=
\ga\bigotimes p^{(\mp)}_j+\frac{1}{2}(1\pm\ga)\bigotimes(p^{(+)}_j- p^{(-)}_j) 
\;\mbox{ for }\;\kappa_j=0 \;,\; p^{(+)}_j\,_<\hspace{-2.4mm}^>\;p^{(-)}_j\;.
\label{P50}
\ee
Inserting now \re{P5} and \re{P50} into 
$P^{(+)}-P^{(-)}=\sum_j (P^{(+)}_j-P^{(-)}_j)$ it 
becomes obvious that for a nonvanishing index of $D^{(+)}$ the extra term in 
\re{P50} prevents $P^{(+)}-P^{(-)}$ from getting the form 
$\ga\bigotimes\Id_s$~, i.e.~that we have
\be
P^{(+)}-P^{(-)}=\ga\bigotimes\Id_s\mb{iff}\mbox{ index }D^{(+)}=0\;.
\label{P5g}
\ee
Thus the naive dealing with $\ga$ which effectively uses 
$\ga\bigotimes \Id_s$ instead of $P^{(+)}-P^{(-)}$ turns out to be only valid 
if the index of $D^{(+)}$ vanishes.

\subsection{Comparison of concepts} \label{comc}

\hspace{3mm}
Clearly the settings to be compared are quite different. In the Atiyah-Singer 
case one considers a differential operator and the concept is that of classical 
fields, while in the lattice approach a subtle limit of a discrete operator 
realizes the quantum concept. 

The basic structural difference is that in the Atiyah-Singer framework a 
nonvanishing index results from the fact that the dimensions of $E^{(+)}_j$ 
and $E^{(-)}_j$ for $\kappa_j=0$ are different while those for all $\kappa_j>0$ 
are the same. In other words, it stems from different cardinalities of the 
spaces $E^{(+)}$ and $E^{(-)}$, which in turn are solely caused by different 
dimensions of the chiral eigenspaces at $\kappa_j=0$. 

In contrast to this in nonperturbative quantized theory no such asymmetry of 
space exists. A nonvanishing index there results from a chirally noninvarinant
part of the action which is conceptually necessary (to avoid doublers). The 
mechanism, reflected by the sum rule \re{res}, then is that nonvanishing of 
the index necessarily implies the existence of a corresponding difference at 
other eigenvalues.

In the Atiyah-Singer case the space structure itself depends on $D\A$, i.e.~on
the particular gauge field. In fact, to 
determine the subspaces $E^{(+)}$ and $E^{(-)}$ which make $E$ up, one has 
first to solve the respective eigenvalue problems. Obviously this has to be 
done for each gauge-field configuration and one has to allow for different 
structures in each case.

On the other hand, with the nonperturbative definition of the quantized theory 
the space structure is fixed and does not depend in any way on $D$ and on the 
gauge field. 

A further difference is that, while the formulation of the quantized theory 
given here is in ${\bf R^4}$, in the Atiyah-Singer case the fields live on
a compact manifold$\,$\footnote{In \cite{ja77} an extension to ${\bf R^4}$ has 
been proposed.  It is, however, an open question wether the proof of 
\cite{at68} can be extended.}.

\section{Analysis of continuum approaches} \label{Comp}\setcounter{equation}{0} 

\subsection{General observations} \label{gam}

\hspace{3mm}
In conventional continuum approaches to quantized theory the form
$D=\sum_{\mu}\gamma_{\mu}D_{\mu}$ of the Dirac operator is used. For this form 
because of $\{\ga,D\}=0$ the anomaly term in the identity \re{gwa2} vanishes. 
Further, since with  hermitean $\gamma$-matrices the spectrum is on the 
imaginary axis according to \re{gDc} only the discrete point at zero 
contributes to the index term. Thus the identity \re{gwa2} degenerates to
\be
\mbox{Tr}\,\ga=I(0)=0 \;.
\label{s0}
\ee
For anti-hermitean $\gamma$-matrices$\,$\footnote{Instead of the antihermitean
operator $D$ with hermitean $\gamma$-matrices $\gamma_{\mu}$ one may consider 
the hermitean operator $D^{\scriptsize a}=
\sum_{\mu}\gamma^{\scriptsize a}_{\mu} D_{\mu}=iD$ with 
antihermitean $\gamma$-matrices $\gamma^{\scriptsize a}_{\mu}=
i\gamma_{\mu}$.\label{ftd}}, with the spectrum on the real axis and 
purely-imaginary-mode rules$^{\ref{ft}}$, again \re{s0} is obtained.

Actually the result \re{s0} is no surprise since proper nonperturbative 
definition of the quantized theory requires discretization also of the action. 
Then the choice $D=\sum_{\mu}\gamma_{\mu}D_{\mu}$ is not appropriate because it
suffers from the doubling phenomenon and the vanishing of anomaly and index is 
exactly what in that case is to be expected.

One should also note that with a function $f(D^2)$ where $f(0)=1$ using
\re{gDc} one gets Tr$(\ga f(D^2))=I(0)$. However, with \re{s0} this gives
\be
\mbox{Tr}\Big(\ga f(D^2)\Big)=I(0)=0 
\label{s1}
\ee
so that the introduction of such a function is seen not to change the result.

In conventional continuum approaches to quantized theory instead of \re{t5ga}
of the Atiyah-Singer case one uses \re{t5g} with the global chiral 
transformation. Such replacement of $P^{(+)}-P^{(-)}$ by $\ga$, however, 
according to \re{P5g} requires a vanishing index. Thus also from the 
Atiyah-Singer point of view \re{s0} and \re{s1} hold.

\subsection{Perturbation theory} \label{comb}

\hspace{3mm}
The question now is how in conventional continuum approaches nevertheless the 
chiral anomaly can arise. In perturbation theory at the level of the Ward 
identity (in the well known triangle diagram) one gets an ambiguity which, if 
fixed in a gauge-invariant way, produces the anomaly term \cite{ad69,bj69}. The 
point is that this fixing of the ambiguity constitutes a chirally noninvariant 
modification of the theory. 

Thus there is no contradiction to the nonperturbative approach. One observes
that while in continuum theory a modification occurs only at the level of the 
Ward identity and is put in by hand, in the nonperturbative theory based on 
the lattice it is already built in into the action and thus is included in the 
theory from the start.

\subsection{Pauli-Villars approach} \label{compv}

\hspace{3mm}
If in continuum perturbation theory the Pauli-Villars (PV) regularization 
is used, in the PV difference ambiguous contributions, being mass-independent, 
drop out so that the PV term gives the anomaly \cite{ad69,bj69}. This has 
suggested a nonperturbative interpretation of it based on an evalution of
$ - \lim_{m\ra\infty}\mbox{Tr}(\ga m(D+m)^{-1})$
as given in \cite{br77}, which neglecting higher orders in the PV mass 
arrives at the desired result. If nonperturbatively correct this would be in 
contradiction to the fact that the anomaly term for 
$D=\sum_{\mu}\gamma_{\mu}D_{\mu}$ vanishes. 

Using $\{\ga,D\}=0$ one sees that $\mbox{Tr}(\ga m(D+m)^{-1})=
\mbox{Tr}(\ga m^2(m^2-D^2)^{-1})$ and further that this expression is 
independent of $m$. For $m\ra0$ it obviously gives the index $I(0)$. Actually
it is just \re{s1} with the choice $f(D^2)=m^2(m^2-D^2)^{-1}$ and therefore 
vanishes.

To check what has been done in \cite{br77} we decompose $D^2$ as 
$D^2=\check{L}+\V$ into parts with and without $\gamma$-matrices, 
\be
\check{L}=\sum_{\mu}D_{\mu}^2 \mb{,} 
\V=\frac{1}{2}\sum_{\mu\ne\nu}\gamma_{\mu}\gamma_{\nu}[D_{\mu}D_{\nu}] \;.
\label{z0} 
\ee
Because at least four $\gamma$-matrices are needed to contribute to the
trace one can write
\be
-\mbox{Tr}\Big(\ga m(D+m)^{-1}\Big)=
     m^2\mbox{Tr}\Big(\ga \g\V\g\V\g\Big)-
     m^2\mbox{Tr}\Big(\ga \g\V\g\V\g\V(D^2-m^2)^{-1}\Big)
\label{pv2}
\ee
where $\g=(\check{L}-m^2)^{-1}$. In \cite{br77} only the term 
$m^2\mbox{Tr}\Big(\ga \g\V\g\V\g\Big)$ of \re{pv2} is kept and evaluating it 
gives the desired result. 

The problem is, however, that neglecting the rest is not allowed because
\be
m^2\mbox{Tr}\Big(\ga \g\V\g\V\g\V(D^2-m^2)^{-1}\Big)
=m^2\mbox{Tr}\Big(\ga \g\V\g\V\g\Big)
\label{zz}
\ee
holds, which reflects the fact that the correct nonperturbative result is 
zero. Thus neglegting the respective term is actually a modification of the 
theory at the level of the Ward identity which is equivalent to what is
done in perturbation theory.

\subsection{Path-integral approach} \label{compi}

\hspace{3mm}
Using the formal path integrals of continuum theory in \cite{fu79} the chiral
anomaly has been claimed to arise from the measure. This can be checked with 
the proper definition of those integrals from the lattice. In Section \ref{Alt}
we have shown that alternative transformations allow to transport chirally 
noninvariant terms to the measure contribution. However, no such transformation
has been used and no such term has been in the action in \cite{fu79}. Further 
in Section \ref{gam} we have pointed out that the chiral transformation with 
\re{t5ga} of the Atiyah-Singer case gives a measure contribution in \re{gwa3}. 
However, the transformation in \cite{fu79} has not been of this type. Thus none
of such mechanisms can be working there.

In \cite{fu79} the result of the local chiral transformation is found to be 
ill-defined. The actual reason of this is that, instead of 
$\mbox{Tr}(\ga\hat{e}(n))\equiv\mbox{Tr}(\ga\bigotimes|n\rangle\langle n|)$ 
with proper vectors $|n\rangle$ which we have in \re{loc}, the formal continuum
integrals there lead to $\sum_k\varphi_k(x)\dg\ga\varphi_k(x)=
\mbox{Tr}(\ga\bigotimes|x\rangle\langle x|)$ with generalized vectors 
$|x\rangle$. The obvious problem then is that $|x\rangle\langle x|$ is no 
projection. This could be fixed by any of the known methods of dealing with
generalized vectors. The most appropriate way, however, is to start properly 
from the discrete definition of the integrals in which case no such problem 
arises.

In the case of the global chiral transformation the proposal in \cite{fu79} 
corresponds to the replacement of Tr$\,\ga$ by Tr$(\ga f(D^2))$ with a function 
$f$ which satisfies $f(0)=1$ and has a suitable behavior at infinity. However,
as discussed in Section \ref{gam} in this situation \re{s0} and \re{s1} apply
and the result is zero not only in nonperturbative quantized theory but also
from the point of view of the Atiyah-Singer theorem.

The desired result in \cite{fu79} is nevertheless obtained essentially 
repeating the procedure of \cite{br77}. Instead of $f(D^2)=m^2(m^2-D^2)^{-1}$ 
there in \cite{fu79} this is done with $f(D^2)=\exp(D^2/m^2)$ and the possible
use of other functions is pointed out.  As in \cite{br77} by keeping only the 
appropriate term the anomaly is obtained. Thus again the theory is modified 
at the level of the Ward identity which is equivalent to what is done in 
perturbation theory.

\section*{Acknowledgement}

\hspace{3mm}
I am grateful to Michael M\"uller-Preussker 
and his group for their continuing kind hospitality. 



\begin{thebibliography}{99}

\bibitem{na94}  R. Narayanan and H. Neuberger, 
	      Nucl. Phys. {\bf B412}, 574 (1994). 
\bibitem{na93}  R. Narayanan and H. Neuberger, 
              Phys. Rev. Lett. {\bf 71}, 3251(1993).
\bibitem{na95}  R. Narayanan and H. Neuberger, 
	      Nucl. Phys. {\bf B443}, 305 (1995). 
\bibitem{gi82}  P.H. Ginsparg and K.G. Wilson, 
              Phys. Rev. D {\bf 25}, 2649 (1982).
\bibitem{ha98}  P. Hasenfratz, V. Laliena, and F. Niedermayer, 
              Phys. Lett. B {\bf 427}, 125 (1998). 
\bibitem{lu98}  M. L\"uscher,
              Phys. Lett. B {\bf 428}, 342 (1998). 
\bibitem{ne98}  H. Neuberger, 
              Phys. Lett. B {\bf 417}, 141 (1998); 
              {\it ibid} {\bf 427}, 353 (1998). 
\bibitem{ke99}  W. Kerler, Phys. Lett. B {\bf 470}, 177 (1999).
\bibitem{ad98}  D.H. Adams, hep-lat/9812003  v4.
\bibitem{su98}  H. Suzuki, Prog. Theor. Phys. {\bf 102}, 141 (1999).
\bibitem{ke81}  W. Kerler, 
          Phys. Rev. {\bf D 23}, 2384 (1981); {\it ibid} {\bf 24}, 1595 (1981).
\bibitem{se82}  E. Seiler and I.O. Stamatescu,
                Phys. Rev. D {\bf 25}, 2177 (1982); 
                err. {\it ibid} {\bf 26}, 534 (1982).
\bibitem{br77}  L.S. Brown, R.D. Carlitz, and C. Lee, 
              Phys. Rev. D {\bf 16}, 417 (1977).
\bibitem{ja77} R. Jackiw and C. Rebbi,
              Phys. Rev. D {\bf 16}, 1052 (1977).
\bibitem{fu79} K. Fujikawa, 
               Phys. Rev. Lett. {\bf 42}, 1195 (1979);  
               Phys. Rev. D {\bf 21}, 2848 (1980); 
               err. {\it ibid} {\bf 22}, 1499 (1980). 
\bibitem{fu99}  K. Fujikawa, 
               Phys. Rev. D {\bf 60}, 074505 (1999). 
\bibitem{ch98}  T.-W. Chiu, 
              Phys. Rev. D {\bf 58}, 074511 (1998). 
\bibitem{at68} M.F. Atiyah and I.M. Singer, Ann. of Math. {\bf 87}, 485 (1968).
\bibitem{at68a}M.F. Atiyah and I.M. Singer, Ann. of Math. {\bf 87}, 546 (1968).
\bibitem{ad00}  D.H. Adams, Chin. J. Phys. {\bf 38}, 633 (2000).
\bibitem{ad98a}  D.H. Adams, hep-lat/9812003  v5.
\bibitem{ad00a}  D.H. Adams, hep-lat/0009026.
\bibitem{ad69} S.L. Adler, Phys. Rev. {\bf 177}, 2426 (1969).
\bibitem{bj69} J. Bell and R. Jackiw, Nuovo Cim. {\bf 60A}, 47 (1969).
\bibitem{al85} L. Alvarez-Gaum\'e, Lectures at International School of
               Mathematical Physics, Erice 1985.
\bibitem{be96}R.A. Bertlmann, {\it Anomalies in Quantum Field Theory}
              (Clarendon Press, Oxford 1996).
\bibitem{ch99}  T.-W. Chiu, 
              Phys. Rev. D {\bf 60}, 034503 (1999). 
\bibitem{ne99} H. Neuberger, Phys. Rev. D {\bf 61}, 085015 (2000).
\bibitem{he99} P. Hernandez, K. Jansen, and M. L\"uscher, Nucl. Phys. 
              {\bf B552}, 363 (1999). 
\bibitem{ke84}  W. Kerler, Z. Physik C {\bf22}, 185 (1984).
\bibitem{ka66}  T. Kato, {\it Perturbation theory for linear operators}
              (Springer, Berlin $\cdot$ Heidelberg $\cdot$ New York 1966),
\bibitem{fu00} K. Fujikawa, Nucl. Phys. {\bf B589}, 487 (2000). 
\bibitem{ke00} W. Kerler, hep-lat/0103024.
\bibitem{ch77}  Y. Choquet-Bruhat, C. Dewitt-Morette, and M. Dillard-Bleick,
             {\it Analysis, Manifolds and Physics}
              (North-Holland, Amsterdam $\cdot$ New York $\cdot$ Oxford 1977),
\bibitem{ka81} L.H. Karsten and J. Smit, Nucl. Phys. 
              {\bf B183}, 103 (1981).
\bibitem{fu98}  K. Fujikawa, 
               Nucl. Phys. {\bf B546}, 480 (1999). 
\bibitem{at73} M. Atiyah, R. Bott, and V.K. Patodi, Inventiones Math. 
               {\bf 19}, 279 (1973); err. {\it ibid} {\bf 28}, 277 (1975).
\end{thebibliography}
\end{document}